\documentclass[sigconf]{acmart}
\usepackage{pifont}
\usepackage{circledsteps}
\usepackage{multirow}
\usepackage{color}
\usepackage{xcolor}
\usepackage{subfigure}
\usepackage{amsmath}
\usepackage{amsthm}
\usepackage{algpseudocode}
\usepackage[ruled]{algorithm2e}

\usepackage{stfloats}

\AtBeginDocument{%
  }

\setcopyright{acmlicensed}
\copyrightyear{2018}
\acmYear{2018}
\acmDOI{XXXXXXX.XXXXXXX}

\acmConference[Conference acronym 'XX]{Make sure to enter the correct
  conference title from your rights confirmation emai}{June 03--05,
  2018}{Woodstock, NY}
\acmISBN{978-1-4503-XXXX-X/18/06}

\begin{document}


\title{\textit{DMGNN}: Detecting and Mitigating Backdoor Attacks in Graph Neural Networks}

\author{Hao Sui}
\affiliation{%
  \institution{Nanjing University of Aeronautics and Astronautics}
  \city{Nanjing}
  \country{China}}
  \email{suihao36@nuaa.edu.cn}

\author{Bing Chen}
\affiliation{%
  \institution{Nanjing University of Aeronautics and Astronautics}
  \city{Nanjing}
  \country{China}}
\email{cb_china@nuaa.edu.cn}
\authornote{Corresponding author.}

\author{Jiale Zhang}
\affiliation{%
  \institution{Yangzhou University}
  \city{Yangzhou}
  \country{China}}
\email{jialezhang@yzu.edu.cn}
\authornotemark[1]

\author{Chengcheng Zhu}
\affiliation{%
 \institution{Yangzhou University}
  \city{Yangzhou}
  \country{China}}
\email{MX120220554@stu.yzu.edu.cn}

\author{Di Wu}
\affiliation{%
  \institution{University of Southern Queensland}
    \city{Toowoomba}
  \country{Australia}}
\email{di.wu@unisq.edu.au}

\author{Qinghua Lu}
\affiliation{%
  \institution{Data61, CSIRO}
  \city{Sydney}
  \country{Australia}}
\email{qinghua.lu@data61.csiro.au}

\author{Guodong Long}
\affiliation{%
  \institution{University of Technology Sydney}
  \city{Sydney}
  \country{Australia}}
\email{guodong.long@uts.edu.au}


\renewcommand{\shortauthors}{Sui et al.}

\begin{abstract}
Graph neural networks (GNNs) are widely implemented in various domains due to the powerful representation learning capability for handling graph structure data. However, recent studies have revealed that GNNs are highly susceptible to multiple adversarial attacks. Among these, graph backdoor attacks pose one of the most prominent threats, where attackers cause models to misclassify by learning the backdoored features with injected triggers and modified target labels during the training phase. Based on the features of the triggers, these attacks can be categorized into out-of-distribution (OOD) and in-distribution (ID) graph backdoor attacks, triggers with notable differences from the clean sample feature distributions constitute OOD backdoor attacks, whereas the triggers in ID backdoor attacks are nearly identical to the clean sample feature distributions. Existing methods can successfully defend against OOD backdoor attacks by comparing the feature distribution of triggers and clean samples but fail to mitigate stealthy ID backdoor attacks. Due to the lack of proper supervision signals, the main task accuracy is negatively affected in defending against ID backdoor attacks.
To bridge this gap, we propose \textit{DMGNN} against OOD and ID graph backdoor attacks that can powerfully eliminate the stealthiness to guarantee the defense effectiveness and improve the model performance. Specifically, \textit{DMGNN} can easily identify the hidden ID and OOD triggers via predicting label transitions based on counterfactual explanation. To further filter the diversity of generated explainable graphs and erase the influence of the trigger features, we present a reverse sampling pruning method to screen and discard the triggers directly on the data level.
Extensive experimental evaluations on open graph datasets demonstrate that \textit{DMGNN} far outperforms the state-of-the-art (SOTA) defense methods, reducing the attack success rate to 5\% with almost negligible degradation in model performance (within 3.5\%).

\end{abstract}




\maketitle

\section{Introduction}
Graph-structured data is ubiquitous across many real-world domains, including transaction networks \cite{ref3}, social networks \cite{ref2}, and protein structure graphs \cite{ref1}.
The latent structure information in the graph data is still unavailable.
Graph neural networks (GNNs) \cite{ref4, ref5} possess powerful representation learning capability to deal with node and edge features on graph structure data, which have shown outstanding performance in various tasks such as node classification \cite{ref6}, graph classification \cite{ref51}, and link prediction \cite{ref8}. The superior-performance GNN models rely heavily on large amounts of training data, prompting many developers to use third-party public graph data to train their models in open web environments. Unfortunately, these third-party public graph data commonly contain erroneous information that causes the GNNs to be vulnerable to backdoor attacks and compromise the security of GNN models.


\begin{figure}
	\centering
	\subfigure[OOD backdoor attack]{
		\begin{minipage}[t]{0.485\linewidth}
			\centering			\includegraphics[width=1\linewidth]{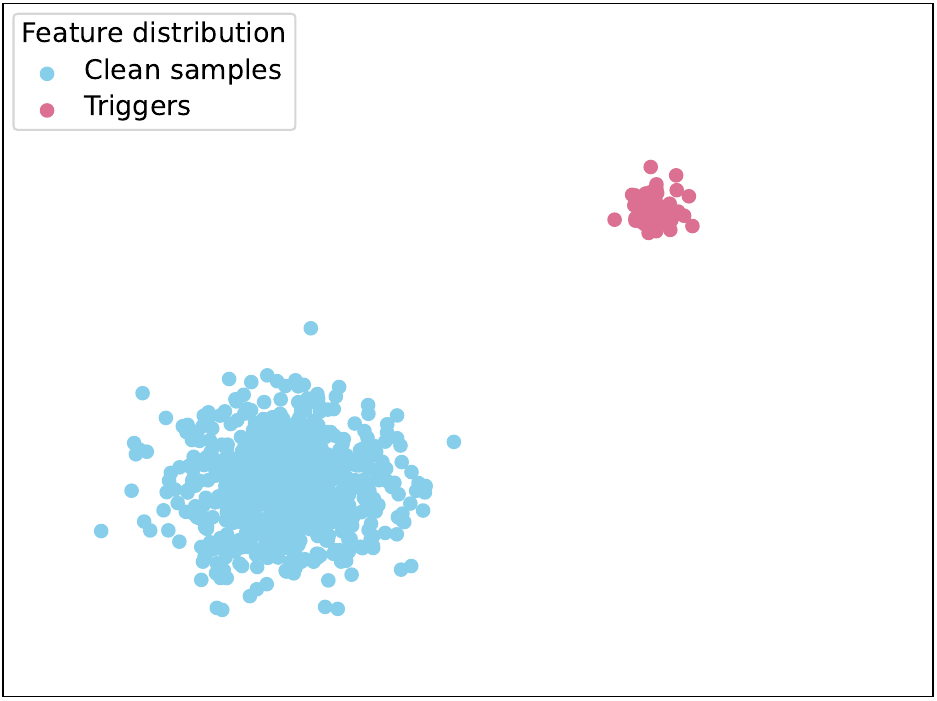}
			\label{F1-1}
	\end{minipage}}
	\subfigure[ID backdoor attack]{
		\begin{minipage}[t]{0.485\linewidth}
			\centering			\includegraphics[width=1\linewidth]{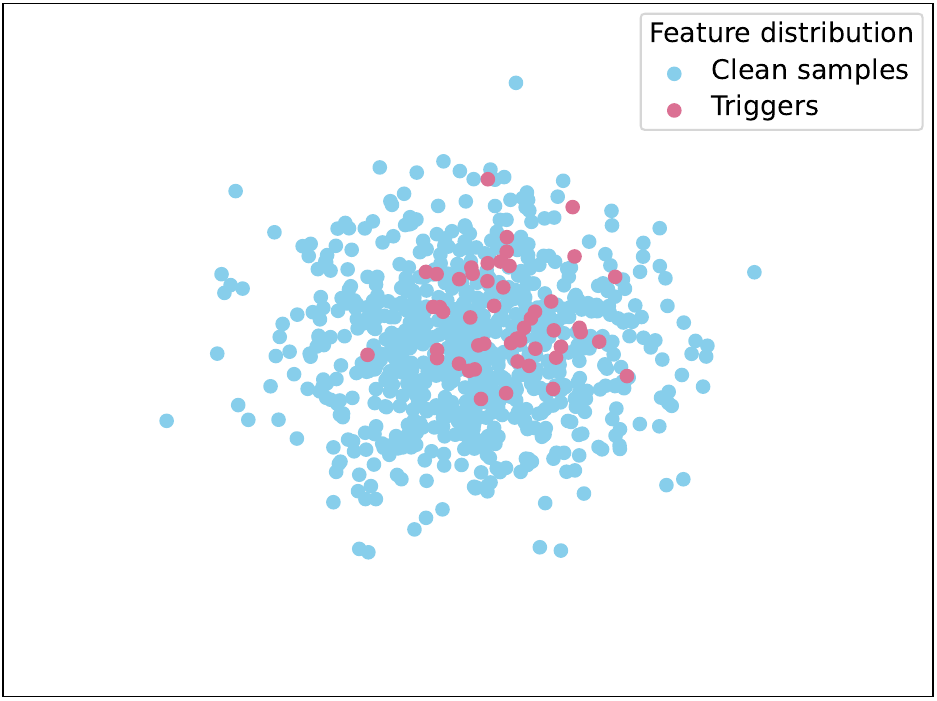}
			\label{F1-2}
	\end{minipage}} 
\caption{Feature distribution in OOD and ID backdoor attacks.}
	\label{F1}
\end{figure}

Backdoor attacks are one of the most serious security threats against GNNs \cite{ref9,ref10,ref11,ref12,ref13,ref14,ref15,ref16,ref17,ref18,ref19,ref20,ref21,ref22,ref23}. The backdoor attacker injects triggers into the original graph and modifies the poisoned sample labels as the target label to train the backdoored model, resulting in triggers and the target label being tightly associated. During testing, samples injected with triggers will be misclassified as the target label, whereas clean samples will be predicted as original labels under the backdoored model. Since the goal of the backdoor attack is to misclassify the backdoored samples,
existing proposals on graph backdoor attacks mainly focus on graph classification and node classification tasks. Attackers design various triggers to increase the attack success rate, which mainly includes modified graph topology structure, tempered node features, and newly generated subgraphs. 
In particular, different types of triggers lead to out-of-distribution (OOD) and in-distribution (ID) backdoor attacks. Figure \ref{F1} fully explains the OOD and ID triggers. The graph backdoor attack with OOD triggers indicates that the distribution of trigger and clean sample features is significantly distinct. In contrast, the features of ID triggers have been incorporated into clean samples and are extremely stealthy in ID graph backdoor attacks.
Therefore, owing to the increase in the variety of attacks and the enhancement of stealthiness, backdoor attacks on GNNs have attracted more attention and concern.

Backdoor defense has been extensively studied in the field of computer vision \cite{ref24,ref25,ref26,ref27,ref28,ref29}. These works can neglect graph topological structure features, which hampers the effectiveness of the defense. To mitigate graph backdoor attacks, prior works have explored strategies mainly consisting of backdoor detection \cite{ref30,ref31} and purification \cite{ref32,ref33,ref34,ref35}. The former approaches focus on distinguishing between backdoored samples and clean samples by leveraging explainable mechanisms. However, these methods only identify the existence of backdoors in samples and fail to effectively eliminate backdoors, making it difficult to mitigate the harm caused by backdoor attacks. Backdoor purification methods \cite{ref32,ref33,ref34} aim to erase the impact of backdoor triggers by distinguishing the backdoor features from benign ones, which can only be applied against OOD graph backdoor attacks. 
Recently, Zhang et al. \cite{ref35} leveraged random edge dropping to differentiate poisoned nodes from normal nodes and design a novel robust training to eliminate the influence of triggers. However, since the correct label information of backdoored samples is lacking during the training process, this method cannot guarantee the label of identified malicious nodes are recovered to their original labels. This phenomenon will lead to lower prediction accuracy of the model for that class when the excessive number of poisoned nodes in the same class, thereby affecting the model performance. \textit{Therefore, it is urgent to need a backdoor defense approach for counter graph backdoor attacks with ID or OOD triggers that aim to effectively alleviate the influence of poisoned data on the model while ensuring high model performance.}

\textbf{Our work.} In this paper, we explore a critical issue about designing a novel backdoor defense approach against both OOD and ID backdoor attacks in GNNs. In essence, we are faced with two challenges: \ding{182} \textit{How to identify hidden ID or OOD triggers for eliminating the stealthiness of the attack}? \ding{183} \textit{How to eliminate the impact of triggers to reduce the effectiveness of the attack and ensure the model performance}? To tackle these challenges, we propose the graph backdoor detection and defense method, named \textit{DMGNN}, which aims to eliminate the stealthiness to guarantee the defense effectiveness and improve the
model performance. Regardless of OOD or ID triggers in graph backdoor attacks, the inherent fact that once triggers are injected into target nodes will cause the predicted label to shift to the target class. Thus, to better discover ID and OOD triggers, we present the counterfactual explanation method to directly find hidden triggers by predicting labeling transitions. Meanwhile, the counterfactual explanation method generates corresponding explainable graphs against each sample feature. Nonetheless, a single sample may have multiple explanation graphs, indicating that the generated graphs exhibit diversity and the triggers cannot be fully removed. To further erase the influence of the backdoor, \textit{DMGNN} introduces reverse sampling pruning to choose the explainable graph that most closely matches the trigger and directly remove this explainable graph from the backdoored data. It ensures improved main task accuracy while achieving efficient defenses.

In summary, the contributions of our paper are:
\begin{itemize}
	\item We propose a novel backdoor detection and mitigation method, named \textit{DMGNN}, in the context of GNNs. It adopts the counterfactual explanation method to precisely identify potential OOD or ID backdoor triggers by analyzing changes in the results of model predictions.
	\item To further mitigate the influence of the triggers due to the diversity of generated counterfactual explanation graphs, we design a reverse sampling pruning method to select and remove the triggers directly on the data level. It can successfully defend OOD and ID backdoor attacks and simultaneously guarantee high model performance.
 
	\item We perform extensive experiments of \textit{DMGNN} on four real-world graph datasets. Experimental results demonstrate the potential of \textit{DMGNN} against three state-of-the-art graph backdoor attacks significantly surpasses the performance of all baseline methods. 
\end{itemize} 

The remainder of this paper is organized as follows. In Section \ref{sec: Background and Related Work}, we discuss the background and related works. In Section \ref{sec: Preliminary Analysis}, we describe the problem definition and the threat model. In Section \ref{sec: Proposed Defense Method}, we introduce our proposed \textit{DMGNN} method. Section \ref{sec: Performance Evaluation} demonstrates the performance evaluation results. Finally, Section \ref{sec: SUMMARY AND FUTURE WORK} summarizes this paper and discusses future works. 

\section{Background and Related Work}
\label{sec: Background and Related Work}
\subsection{Backdoor Attack and Defense on DNNs}
\label{sec: Backdoor Attack and Defense}
Backdoor attacks are a powerful type of attack in deep neural networks (DNNs), where the attacker either manipulates the original data or tampers with the model parameters to produce a poisoned model, significantly compromising the accuracy of the model's predictions. To increase the effectiveness of the attack, the first backdoor attack on DNNs was proposed by Gu et al.\cite{ref38} via injecting pixel patches as triggers into the raw data. Based on this, several studies started to design imperceptible triggers. Saha et al. \cite{ref39} utilized the hidden location of the triggers to improve the success rate of the attack. In addition, Tao et al. \cite{ref40} designed a new distribution-preserving backdoor attack capable of consistent backdoor features with the normal features distribution for better deception modeling. The above methods are mainly for centralized scenarios, in order to fill the gap of distributed backdoor attacks, Naseri et al. \cite{ref41} designed a backdoor attack in federated learning.

In response to the above various backdoor attacks, how to eliminate the backdoor has attracted the attention of many researchers. To mitigate the harm of backdoors, Wang et al. proposed the NeuralCleanse method \cite{ref24} to identify backdoors by measuring trigger impact and leveraging neuronal activation regions. DBD \cite{ref25} utilized a self-supervised approach to eliminate the effect of tampered labels in the backdoored samples and then purified the backdoored model using fine-tuning. The SAGE method \cite{ref26} employed self-attention distillation to eliminate backdoor features in the deep layers through the shallow information of the model. To enhance the robustness of backdoor defense, RAB \cite{ref27} adopted the smoothing training method to avoid noisy distribution. BayBFed \cite{ref28} reduced the attack success rate against federated learning backdoor attacks that detect and filter malicious updates by employing the hierarchical process. Unfortunately, the method may degrade the model performance. Thus, zhang et al. proposed FLPurifier \cite{ref29} to extract label-independent feature information, which breaks the strong connection between target labels and backdoor features.

\subsection{Backdoor Attack and Defense on GNNs}
\label{sec: Backdoor Attack and Defense on GNNs} 
Backdoor attacks are not only effective in DNN models but also highly aggressive in GNN models. The existing graph backdoor attacks are mainly applied to graph classification and node classification tasks. 

\ding{182}\textbf{Applied to graph classification}. In graph classification tasks, attackers mainly create different hidden triggers to launch backdoor attacks. Zhang et al. \cite{ref16} first proposed a graph backdoor attack by generating fixed triggers to replace the raw subgraphs directly and modifying the graph labels as target labels. Further, studies \cite{ref13, ref14, ref15, ref21} designed adaptive triggers that can generate corresponding triggers based on different sample features, allowing for improved attack success rates. In addition, Xu et al. \cite{ref19} designed a clean-label graph backdoor attack so that the attacker can successfully launch a backdoor attack without modifying the label. While these methods use the generated subgraphs as triggers, there are still some methods that utilize the generated node features as triggers and replace the original node features. Xu et al. \cite{ref10} identified important features in nodes as triggers in explainable manners. Dai et al. \cite{ref11} used semantic information to generate features in nodes as triggers.

\ding{183}\textbf{Applied to node classification}. In node classification tasks, attackers mainly add the newly generated nodes or subgraphs as triggers to the target samples. Then, the attacker modifies the label of the target sample as the target class. GCBA \cite{ref20} proposed a graph backdoor attack for self-supervised scenarios by injecting new nodes into the data to make the data poisoned. UGBA \cite{ref18} designed adaptive subgraphs as triggers selectively inserted into the target nodes for backdoor attacks. Since the trigger features and normal sample features have obvious differences easily recognized, Zhang et al. \cite{ref23} designed the in-distribution graph backdoor attack to keep the generated trigger and normal sample features consistent, which improves the stealthiness of the attack. 

To mitigate the harm caused by graph backdoor attacks, current works have explored some strategies, categorized into two types: \ding{182} \textit{Backdoor detection}: Some research \cite{ref30,ref31} focused on distinguishing between backdoor samples and clean samples by leveraging explainable mechanisms. \ding{183} \textit{Backdoor purification}: Most purify methods \cite{ref32,ref33,ref34} identified and filtered backdoor features by the difference between benign and backdoor sample features. These methods presupposed that they could only be applied against OOD graph backdoor attacks. To handle ID graph backdoor attacks, Zhang et al. \cite{ref35} leveraged random edge dropping to distinguish poisoned nodes from normal nodes and design a novel robust training to eliminate the influence of triggers. Nevertheless, this method cannot guarantee that the identified malicious nodes are recovered to their original labels, due to the lack of supervised training. If the excessive number of poisoned nodes in the same class, it will lead to lower prediction accuracy of the model for its class, thus affecting the model performance.

\subsection{Explainability on Graph Neural Network}
\label{sec: Explainability on Graph Neural Network}
Graph neural networks are widely used in various fields \cite{ref1, ref2, ref3}. Explainable methods in GNNs \cite{ref42, ref43, ref44, ref45, ref46, ref47} have been proposed to better comprehend the model of the prediction mechanism. The purpose of the graph explainable technique is to provide the user with an understanding of the potential relationship between the input samples and the predicted results. Based on the difference in explanation results, graph explanation techniques might be categorized into instance level \cite{ref42, ref43, ref44, ref45, ref46} and model level \cite{ref47}. The instance level is the explanation that provides features for each input graph. Whereas, the model level is the explanation of the model globally without considering the input graph features. Owing to the adaptive property of triggers, we employ the instance-level explainable approach in this paper. Among the instance-level methods, the most typical one is the GNNExplainer method \cite{ref44}. It was a perturbation-based method utilizing optimization techniques to learn soft masks of edge and node features for identifying the vital explainable edges and node features of the graph.

\section{Preliminary Analysis}
\label{sec: Preliminary Analysis}
\subsection{Problem Definition}
Our preliminary introduction shows the relevant definitions for graphs, including the fundamental theory of graphs, graph neural networks applied in node classification tasks, and the vulnerability of GNNs against backdoor attacks.

\textbf{Graph}. 
The graph consists of a series of nodes and edges. It is widely researched owing to its special topology structure. A graph can be defined as $\mathcal{G}$ = ($\mathcal{V}$, $\mathcal{E}$, $\mathcal{X}$, $\mathcal{Y}$), where $\mathcal{V} = \lbrace v_{1},v_{2},...,v_{\mathcal{N}}\rbrace$ denotes each node in the graph and $v_{i}$ denotes the $i$-th node. $\mathcal{E} = \lbrace e_{1},e_{2},...,e_{\mathcal{M}}\rbrace$ denotes the collection of edges and $e_{i}$ denotes the $i$-th edge. $\mathcal{X}$ $\in \mathbb{R}^{|V| \times S}$ indicates each node features, where $\mathcal{S}$ represents the feature dimension of each node. $\mathbb{R}$ denotes the matrix of size ${|V| \times S}$. $\mathcal{Y}$ denotes the label of samples. We denote the topology structure of the graph by $A=\{0,1\}^{n\times n}$, which is $A_{ij}=1$ if an edge exists between nodes $i$ and $j$ and $A_{ij}=0$ if it does not exist. In graph learning, nodes mainly learn node features and topology structure information.

\textbf{Node classification task}. The main components of the graph learning task are node classification and graph classification. Node classification is to classify each node and graph classification is to classify each graph sample. In this paper, we mainly explore node classification and give a detailed description. In the node classification task, the graph dataset can be defined as $Dataset=\{(v_1,y_1),(v_2,y_2),...,(v_\mathcal{N},y_\mathcal{N})\}$, where $v_i$ represents the $i$-th node and $y_i$ represents the label of the $i$-th node. The node features are input into the GNN model for training and the classifier is leveraged to predict the label of this node. The performance of the model is judged by comparing the consistency of the node prediction with the true label.

\textbf{Graph Backdoor attack}. 
The attacker injects a designed trigger into the original dataset $D$ to construct a backdoor dataset $D_d$, which is then used to train a GNN model $\mathcal{G}_o$. The trained model becomes a backdoor model $\mathcal{G}_d$. When the user inputs test data into the backdoor model, test data with the injected backdoor $n_d$ will be misclassified as the target class $y_d$, whereas clean test data $n_o$ will be classified as the original class $y_o$. The attacker needs to ensure the effectiveness of the attack on the poisoned data and the evasiveness of the attack on the clean data. In conclusion, the objectives of the attacker can be outlined as follows:

\begin{equation}
\left.\left\{\begin{array}{l}\mathcal{G}_d(n_d)=y_d\\
\mathcal{G}_d(n_o)=\mathcal{G}_o(n_o)=y_o\end{array}\right.\right.
\end{equation}

\begin{figure*}
	\centering
	\includegraphics[width=1\linewidth]{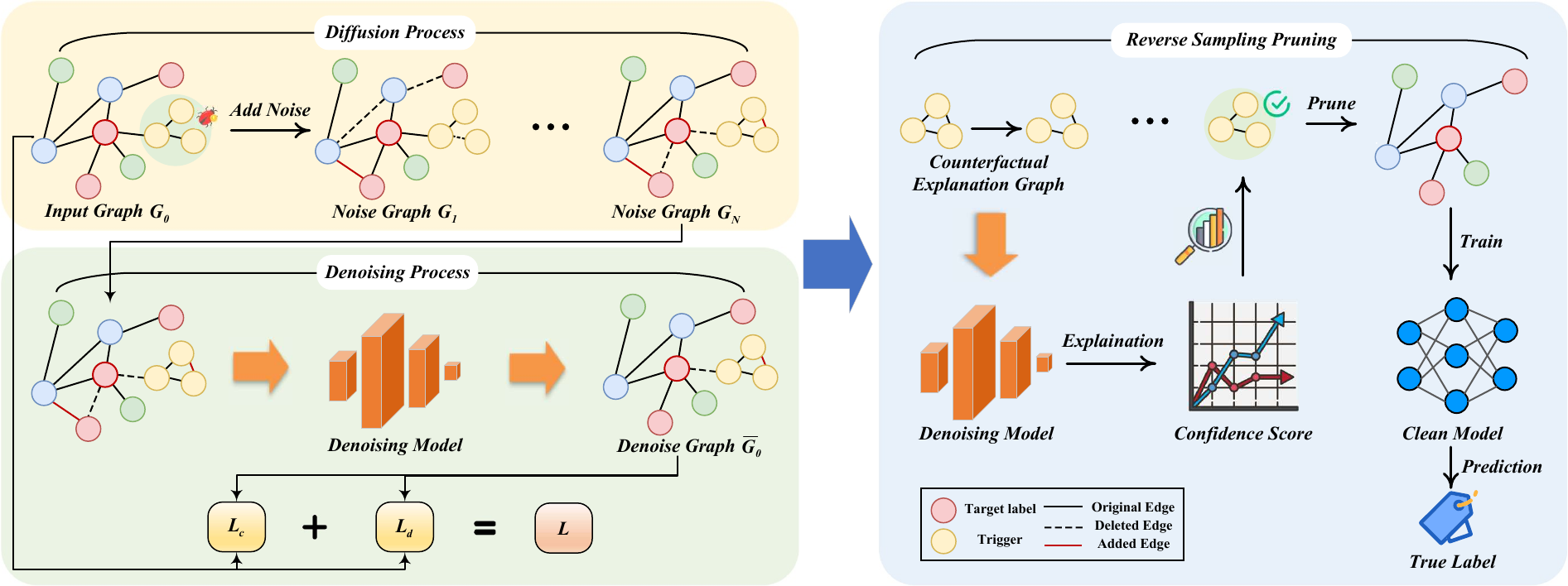}
	\caption{
 The overview of \textit{DMGNN}. In \textit{Stage 1}, identify the hidden ID or OOD triggers by diffusion and denoising processes. \textit{Stage 2}, further eliminate triggers via reverse sampling pruning in the data level.
 }
	\label{F2}
\end{figure*}

\subsection{Threat Model}
\subsubsection{Attacker's Goal and Capability}
The attacker aims to improve the effectiveness and stealthiness of backdoor attacks in GNNs.

\noindent \textbf{Effectiveness}: To enhance the effectiveness of the graph backdoor attack, the attackers create different triggers (nodes or subgraphs) and add these triggers to target nodes. The hidden ID and OOD triggers that are created can boost the aggressiveness of backdoor models, resulting in test samples being predicted to be in the target class once the hidden triggers are added.

\noindent \textbf{Stealthiness}: To enhance the stealthiness of the graph backdoor attack, the attacks need to maintain a high model classification accuracy on clean test samples.

In response to the purpose of the above attack, the attacker trains the loss as follows:

\begin{equation}
Loss=\alpha\mathcal{L}(\mathcal{G}(G_d),y_d)+(1-\alpha)\mathcal{L}(\mathcal{G}(G_o),y_o)
\end{equation}

In the \textit{DMGNN} setup, the attacker inserts ID or OOD triggers on the target node and changes the target node label to increase the association of the trigger with the target class. Therefore, the attacker only knows the training data information and has no knowledge of the model information.

\subsubsection{Defender's Goal and Capability}
The goal of defenders is to reduce the success rate of backdoor attacks while ensuring that model performance is not affected. Existing defense methods only eliminate the effect of OOD triggers, \textit{DMGNN} can effectively defend against graph backdoor attacks with ID or OOD triggers. The defender's capability is that they can only get access to existing third-party public data, but they don't know which data has been injected with a backdoor. Meanwhile, for multi-class datasets, the defenders do not know the target class of the backdoor attack, which adds difficulty and challenge to the defense.

\section{Proposed Defense Method}
\label{sec: Proposed Defense Method}
\subsection{Overview}
In this section, we provide a comprehensive depiction of \textit{DMGNN}, which aims to defend against graph backdoor attacks with ID and OOD triggers. With this comes two challenges: \ding{182} \textit{How to identify the hidden ID or OOD triggers against the graph backdoor attac}k; \ding{183} \textit{How to efficiently eliminate triggers while ensuring model performance}; To tackle the above challenges, a novel overview of \textit{DMGNN} is presented, which is illustrated in Figure \ref{F2}. \textit{DMGNN} implementation consists of two primary modules: \textit{Identify Trigger} and \textit{Eliminate Trigger}. To solve the first challenge, we present the counterfactual explanation method to effectively identify hidden triggers in the \textit{Identify Trigger} module. The method exploits the property that backdoored samples injected into triggers will change their predictions as the target class on the backdoored model. The counterfactual explanation is to gain the minimum perturbation graphs that predict sample label overturning and perfectly match the properties of the triggers, thus identifying potential triggers. To solve the second challenge, \textit{DMGNN} leverages reverse sampling pruning to remove the impact of the trigger on the backdoored data in the \textit{Eliminate Trigger} module. Since the generated counterfactual explainable graphs have diversity and are unable to eliminate triggers accurately.

\subsection{Identify Trigger via Counterfactual Explanation}
\label{sec: Identify Trigger via Counterfactual Explanation}
The crucial to the success of a graph backdoor attack is to make the model misclassified backdoor samples with triggers as the target class. It matters not whether the features of the trigger are similar to those of the normal samples or cause inconsistent feature distributions. Therefore, we only need to find the perturbation graph that enables the backdoor samples to generate the target predicted label, as well as guarantee that the model will predict it as the original label when that perturbation graph is removed. This perturbation graph is the trigger that leads to the key factors for backdoor success.

\textit{DMGNN} leverages the counterfactual explanation method to generate the perturbation graph that will cause the change in predicted results. The counterfactual explanation operations include diffusion and denoising processes. We first add noise to the original input graph by forward diffusion. Let $N$ as the noise level parameter. With the increase of noise level, the forward diffusion process will generate a series of noise graphs $\{G_{1}, \cdots, G_{N}\}$ for providing the more comprehensive counterfactual explanation graphs. At the noise level $n$, the adjacency matrix of the graph represents $A_{n}$, where the $ij$-th element is denoted $a_n^{ij}$ in the adjacency matrix. The diffusion process is formulated as follows:

\begin{equation}
\label{Eq3}
q(G_n|G_{n-1})=\prod_{ij}q(\boldsymbol{a}_n^{ij}|\boldsymbol{a}_{n-1}^{ij}), q(G_n|G_0)=\prod_{ij}q(\boldsymbol{a}_n^{ij}|\boldsymbol{a}_0^{ij})
\end{equation}

\noindent where $G_{0}$ denotes the original input graph, $G_{n}$ denotes the added noise graph at the noise level $n$. $q(\boldsymbol{a}_n^{ij}|\boldsymbol{a}_{n-1}^{ij})$ represents the diffusion process for each element, which can be written as $\mathrm{Cat}(\boldsymbol{a}_n^{ij};\boldsymbol{D}=a_{n-1}^{ij}Q_n$. $\boldsymbol{Q}_n\in\mathbb{R}^{2\times2}$ is a transition matrix and $\mathrm{Cat}(x;P)$ represents a class distribution with probability vector $P$ of the vector $x$.

\textit{DMGNN} further performs the denoising process to generate counterfactual explanation graphs that are very similar to the original input graph. We input the original graph $G_0$, the node features of the original graph $X_0$, and the noise adjacency matrix corresponding to the noise graph $G_n$ into the denoising model $p(G_0|G_n)$. The denoising model utilizes a powerful graph network to transform the noisy graph as a denoised dense adjacency matrix \cite{ref36}. Since the dense graph represented by the dense adjacency matrix is not representative of the explainable graph, we use resampling methods to convert the dense adjacency matrix into a sparse adjacency matrix and an explainable graph relative to it \cite{ref37}.

In particular, counterfactual explanation involves selecting a class and transforming it into the other classes to obtain the minimally perturbed graph. However, the initially selected class is not necessarily the target class. To effectively identify triggers, it is crucial first explicitly to find the target class. If the selected class is the target class, the counterfactual explanation graph obtained from the other class samples will consistently flip their labels to the target class. Conversely, if the selected class is not the target class, applying the counterfactual explanation graph added to other class samples may not cause a shift to the target class, as important features also exist in other classes.

To identify more realistic triggers with in-distribution, the counterfactual explanation method must ensure that the generated graph not only causes a shift in the prediction result but also remains close to the original graph features. Thus, \textit{DMGNN} design an exclusive loss function to handle the demand for generated counterfactual explainable graphs. We leverage the counterfactual loss $\mathcal{L}_{c}$ to optimize the performance of the model counterfactual, as follows:

\begin{equation}
\label{Eq4}
\mathcal{L}_{c}=-\mathbb{E}_{q(G_0)}\mathbb{E}_{n\sim[0,N]}\mathbb{E}_{q(G_n|G_0)}\mathbb{E}_{p(\overline{G}_0|G_n)}\log\left(1-f(\overline{G}_0)[\hat{Y}_{G_0}]\right)
\end{equation}

\noindent where $f$ represents the GNN classifier, $\overline{G}_0$ denotes the denoised graph and $\hat{Y}_{G_0}$ represents the raw label of $G_0$, $q(G_0)$ denotes the distribution of $\overline{G}_0$, $f(\overline{G}_0)[\hat{Y}_{G_0}]$ is the probability that the denoised graph $\overline{G}_0$ is predicted to be $\hat{Y}_{G_0}$ under the GNN classifier $f$. 

To better match the in-distribution triggers in graph backdoor attack, we set the distribution loss $\mathcal{L}_{d}$ to optimize the in-distribution property. Therefore, we leverage a re-weighted strategy that the evidence lower bound rebuilds the original distribution on the negative log-likelihood. Formally, the $\mathcal{L}_{d}$ can be calculated as follows:

\begin{equation}
\label{Eq5}
\mathcal{L}_{d}=-\mathbb{E}_{q(G_0)}\sum_{n=1}^N\left(1-2\cdot\bar{\beta}_n+\frac{1}{N}\right)\mathbb{E}_{q(G_n|G_0)}\log D\left(G_0\mid G_n\right)
\end{equation}

\noindent where $\bar{\beta}_n$ denotes the probability of shift. The goal is to ensure that the original graph $G_0$ and the denoised counterfactual explanation graph $D\left(G_0\mid G_n\right)$ are close in terms of feature distribution, achieving in-distribution property. Considering that are simultaneously consistent with counterfactual and in-distribution properties, the loss of \textit{DMGNN} is a combination of the counterfactual loss and the distribution loss: 

\begin{equation}
\mathcal{L} = \lambda\mathcal{L}_{c}+\mathcal{L}_{d}
\end{equation}

\noindent where $\lambda$ denotes a hyper-parameter for weighing counterfactual and in-distribution properties. To explore the effect of this hyper-parameter on the defense effect, we set up a series of different $\lambda$ for analysis in the experience. This manipulation avoids the impact of the feature distribution of the explanation graph on excessively large or small counterfactual properties. If hyper-parameter $\lambda$ is too large, it may generate out-of-distribution features, even if it can easily alter the model's predicted label. The process of identifying triggers via counterfactual explanation is shown in Algorithm \ref{A1}.

\subsection{Eliminate Trigger via Reverse Sampling Pruning}
\label{sec: Eliminate Trigger via Reverse Sampling Pruning}
\textit{DMGNN} captures potential counterfactual explainable graphs against each sample by adding noise and removing counterfactual irrelevant edges. The counterfactual explainable graph conforms to a similar distribution to the other sample features under the constraint of $\mathcal{L}_{d}$. However, the diversity of the generated counterfactual explainable graphs can provide multiple explainable graphs, and all of these explainable graphs can guarantee the robustness of the counterfactual explanation under noise.

Since the generated ID counterfactual explainable graphs are diverse and fail to eliminate triggers completely, \textit{DMGNN} further eliminates triggers by reverse sampling pruning of the recognized diversity counterfactual explanation graphs. We set the target class as $T$ found in Section \ref{sec: Identify Trigger via Counterfactual Explanation}. The reverse sampling step can be defined as follows:

\begin{equation}
q_{T}(g_{n-1}^{r}|g_{n}^{r})\propto D(\overline{G}_{0}|g_{n}^{r})q(g_{n-1}^{r}|\overline{G}_{0})f(T|\overline{G}_{0})
\end{equation}

\noindent where $g_{n}^{r}$ represents the counterfactual explanation graphs. $g_{n-1}^{r}$ indicates counterfactual explanation graphs after one reverse sampling. Our goal is to ensure that the counterfactual explanation graphs are highly explainable for the target class while satisfying the highest prediction confidence that the poisoned nodes with the trigger are removed are predicted to be in the non-target class, thus improving the model performance.

To make the explainable graphs approximate the triggers, we repeat the above steps in order to increase the explainable confidence about the target class. We set the maximum explainable confidence threshold for the target class to $p$. When the explainable confidence reaches this threshold $p$, we further filter the obtained explainable graphs in order to improve the defense success rate. Therefore, \textit{DMGNN} will prune the explainable graph that computes the highest predictive confidence for the non-target classes. Since the loss function of $\mathcal{L}_{c}$ is used to change the prediction labels, the model explanation process only requires the use of the distribution loss $\mathcal{L}_{d}$. Other operations are consistent with the denoising model training as Equation 3. For the explanation graphs, We first set the number $A$ nodes, one-half edge probability of edges to random sample by an \text{Erdős-Rényi} graph. Secondly, we choose the graph processed by the denoising model to compute the explainable confidence score for the target label. Finally, we eventually obtained the explainable graphs that most closely match the triggers for each poisoned node. \textit{DMGNN} will prune the explainable graphs for each poisoned node thus eliminating the effect of the hidden triggers accurately.

\section{Performance Evaluation}
\label{sec: Performance Evaluation}
In this section, we will evaluate the effectiveness of our proposed \textit{DMGNN} on four real-world datasets against three SOTA graph backdoor attack methods, we will answer three key research questions:

\begin{itemize}
    \item {\Circled{\textbf{RQ1}}} Superiority of \textit{DMGNN}: Can our proposed method be superior compared to the three defense baseline methods against the three SOTA graph backdoor attacks?
	
 \item {\Circled{\textbf{RQ2}}} Effectiveness of \textit{DMGNN}:
    What is the effectiveness of \textit{DMGNN} in other different settings?

    \item {\Circled{\textbf{RQ3}}} Completeness of \textit{DMGNN}:
    What is the importance of the components of \textit{DMGNN} perform?
\end{itemize}

\subsection{Experimental Settings}
\label{sec: Experimental Settings}

\subsubsection{Dataset}
\label{sec: Dataset}
We leverage four public datasets to evaluate the performance of the present method on the node classification task, including Cora, Pubmed \cite{ref48}, Flickr \cite{ref49}, and OGB-arxiv \cite{ref50}. The specific information of these four datasets is shown in detail in Table \ref{table: Analysis of the datasets} and Appendix \ref{sec: Details on Datasets}.

\subsubsection{Attack Methods}
To verify the superiority of the present method, we select three SOTA attack methods under the node classification task: GTA \cite{ref13}, UGBA \cite{ref18}, and DPGBA \cite{ref40}. In particular, GTA can be applied to both node classification and graph classification, while we will consider only the node classification task in this paper. (\textit{i}) GTA injects the generated adaptive triggers into the original nodes for graph backdoor attack in the node classification task. It mainly leverages node features and topology structure to generate adaptive triggers; (\textit{ii}) UGBA selectively injects triggers and target labels in poisoned samples to achieve unnoticeable backdoor attacks; (\textit{iii}) DPGBA is an ID backdoor attack can match trigger features with normal sample feature distribution. The attack is difficult to detect only by comparing the variability of trigger features.

\subsubsection{Baseline Defense Methods}
\label{sec: Baseline Defense Methods}
To demonstrate the superiority of our method, we choose three baseline defense methods for comparison, including Prune, Prune+LD \cite{ref18}, and RIGBD \cite{ref35}. 
(\textit{i}) Prune performs strategic pruning by cosine similarity. Edges connected by nodes with low cosine similarity are pruned; (\textit{ii}) Prune+LD not only cuts off the edges but also discards the labels of the nodes connected to the edges, thus reducing the effect of poisoned nodes on malicious labels; (\textit{iii}) RIGBD utilizes random edge dropping to find poisoned nodes and designs a new robust training strategy to defend the effects of triggers.

\subsubsection{Evaluation Metrics}
We adopt the attack success rate (\emph{ASR}) and the accuracy (\emph{ACC}) metrics to evaluate the attack effectiveness of \textit{DMGNN}. (\textit{i}) \emph{ASR} is a metric designed to evaluate the proportion of backdoor samples successfully classified as target labels under the backdoor model; (\textit{ii}) \emph{ACC} measures the classification of clean samples to assess whether clean samples are affected by the backdoor model. A better graph defense backdoor method requires ensuring both a lower ASR and a higher ACC.

\begin{table*}[ht]\small
	\centering
 \caption{\textbf{The defense performance of \textit{DMGNN} compared with the baseline methods}
 \label{table: Compare with the baseline methods}}
  \renewcommand{\arraystretch}{1} 
 \resizebox{\textwidth}{!}{
\begin{tabular}{cl|cl|cccccccccc}
\toprule[1pt]
\multicolumn{2}{c}{}              & \multicolumn{2}{c}{}            & \multicolumn{2}{c}{Before} & \multicolumn{2}{c}{Prune} & \multicolumn{2}{c}{Prune+LD} & \multicolumn{2}{c}{RIGBD} & \multicolumn{2}{c}{\textit{DMGNN}} \\ \cline{5-14} 
\multicolumn{2}{c}{\multirow{-2}{*}{Dataset}}   & \multicolumn{2}{c}{\multirow{-2}{*}{Attack method}} & ASR          & ACC         & ASR         & ACC         & ASR           & ACC          & ASR         & ACC         & ASR                     & ACC                     \\ \hline  \hline
\multicolumn{2}{c}{}                            & \multicolumn{2}{c}{GTA}                             & 0.904        & \textbf{0.823}       & 0.245       & 0.789       & 0.204         & 0.797        & 0.059       & 0.794       & \textbf{0.005}                   & 0.821                   \\
\multicolumn{2}{c}{}                            & \multicolumn{2}{c}{UGBA}                            & 0.943        & \textbf{0.831}       & 0.896       & 0.792       & 0.875         & 0.778        & 0.072       & 0.805       & \textbf{0.011}                   & 0.817                   \\
\multicolumn{2}{c}{\multirow{-3}{*}{Cora}}      & \multicolumn{2}{c}{DPGBA}                           & 0.956        & \textbf{0.827}       & 0.902       & 0.802       & 0.883         & 0.793        & 0.123       & 0.799       & \textbf{0.014}                   & 0.811                   \\ \hline
\multicolumn{2}{c}{}                            & \multicolumn{2}{c}{GTA}                             & 0.855        & \textbf{0.847}       & 0.329       & 0.774       & 0.253         & 0.748        & 0.063       & 0.801       & \textbf{0.009}                   & 0.836                   \\
\multicolumn{2}{c}{}                            & \multicolumn{2}{c}{UGBA}                            & 0.894        & \textbf{0.851}       & 0.879       & 0.814       & 0.831         & 0.802        & 0.055       & 0.782       & \textbf{0.015}                   & 0.824                   \\
\multicolumn{2}{c}{\multirow{-3}{*}{Pubmed}}    & \multicolumn{2}{c}{DPGBA}                           & 0.899        & \textbf{0.853}       & 0.894       & 0.807       & 0.882         & 0.801        & 0.118       & 0.785       & \textbf{0.021}                   & 0.819                   \\ \hline
\multicolumn{2}{c}{}                            & \multicolumn{2}{c}{GTA}                             & 0.879        & 0.451       & 0.09        & 0.449       & 0.07          & 0.403        & 0.079       & 0.419       & \textbf{0.013}                   & \textbf{0.462}                   \\
\multicolumn{2}{c}{}                            & \multicolumn{2}{c}{UGBA}                            & 0.917        & \textbf{0.461}       & 0.804       & 0.431       & 0.853         & 0.423        & 0.087       & 0.403       & \textbf{0.017}                   & 0.449                   \\
\multicolumn{2}{c}{\multirow{-3}{*}{Flickr}}    & \multicolumn{2}{c}{DPGBA}                           & 0.922        & \textbf{0.456}       & 0.872       & 0.405       & 0.856         & 0.411        & 0.128       & 0.408       & \textbf{0.025}                   & 0.438                   \\ \hline
\multicolumn{2}{c}{}                            & \multicolumn{2}{c}{GTA}                             & 0.906        & \textbf{0.676}       & 0.05        & 0.594       & 0.04          & 0.588        & 0.065       & 0.609       & \textbf{0.007}                   & 0.671                   \\
\multicolumn{2}{c}{}                            & \multicolumn{2}{c}{UGBA}                            & 0.954        & \textbf{0.671}       & 0.906       & 0.597       & 0.884         & 0.605        & 0.082       & 0.617       & \textbf{0.009}                   & 0.658                   \\
\multicolumn{2}{c}{\multirow{-3}{*}{OGB-arxiv}} & \multicolumn{2}{c}{DPGBA}                           & 0.949        & \textbf{0.677}       & 0.889       & 0.603       & 0.895         & 0.615        & 0.114       & 0.603       & \textbf{0.013}                   & 0.649                   \\
\toprule[1pt]
\end{tabular}
    }
\end{table*}

\subsubsection{Implementation Details} 
\textit{DMGNN} provides an effective defense against graph backdoor attacks in node classification tasks. In our experiments, we divided each dataset as the training set and testing set with the proportions of 80\% and 20\%. In addition, we set the poisoned data as 3\% of the training set. Specifically, we choose to inject different numbers of backdoor samples in different classes. In the manner of injecting the backdoor, we will strictly follow the settings in the SOTA methods \cite{ref13, ref18, ref40}. The GNN model employs a 2-layer GIN and sets the dimension of each hidden layer to 128. To demonstrate the transferability of the backdoor defense approach, we defend different architectures of GNNs, including GCN and GAT (see Appendix \ref{sec: Different types of GNNs}). In the counterfactual explanation, we set a hyper-parameter of the loss function $\lambda$ as 0.1. The upper explainable confidence $p$ threshold is set to 0.9 in reverse sampling pruning. Besides that, our experiment set the training epochs as 1000 and the learning rate as 0.001. For the fairness of our experimental results, we test and average five experiment results to avoid the instability of only one result.

\subsubsection{Experimental Environment} \textit{DMGNN} is implemented by the PyTorch framework in Python. The environment used in our experiments is Intel(R) Xeon(R) Silver 4310 CPU 377G (CPU), NVIDIA RTX A6000 (GPU), 75GiB memory, and Ubuntu 20.04 (OS).

\subsection{Defense Performance of \textit{DMGNN}  \Circled{\textbf{RQ1}}}
To answer RQ1, we will compare three baseline methods on four real datasets against three SOTA graph backdoor attacks. The three baseline defense methods have been given a thorough description in Section \ref{sec: Baseline Defense Methods} and the defense effectiveness of \textit{DMGNN} is shown in Table \ref{table: Compare with the baseline methods}. 

As can be found from the experimental results in Table \ref{table: Compare with the baseline methods}, the ASR of \textit{DMGNN} is lower than all the baseline defense methods, which shows the superiority of the defense of our method against different graph backdoor attacks. Meanwhile, the ACC of \textit{DMGNN} is higher than the three baseline methods. Although the ACC of \textit{DMGNN} is lower than the ACC of the method without being attacked in most attack methods, the discrepancy between them is less than 3.5\%. The Prune and Prune+LD methods are not satisfactory against UGBA and DPGBA attacks, despite the fact that they can achieve good results under the GTA attack method. In particular, the ASR drop after Prune defense is only within 6\% against the DPGBA attack method. This is because the triggers generated by DPGBA have the same feature distribution as the normal samples, which causes the Prune method to be unable to differentiate the backdoor samples through the feature distribution variance. RIGBD reduces the attack success rate to less than 0.15, indicating that the defense method can handle ID backdoor attacks through random remove edge and robust training. However, the method reduces the ASR by minimizing the prediction confidence of the poisoned nodes to the target class, and cannot guarantee that the poisoned nodes are recovered as the original labels, due to the lack of supervised training. When there are too many poisoned nodes in the same class, it will lead to a decrease in the prediction accuracy of the model for that class and affect the model performance. Therefore, the results of RIGBD are worse than \textit{DMGNN} when the number of injection trigger samples is not consistent in each class. In addition, we also validate the impact of \textit{DMGNN} under different model architectures (e.g., GCN, GAT), which is described in detail in Appendix \ref{sec: Different types of GNNs}.

\begin{figure*}
	\centering
	\subfigure[GTA]{
		\begin{minipage}[t]{0.325\linewidth}
			\centering			\includegraphics[width=1\linewidth]{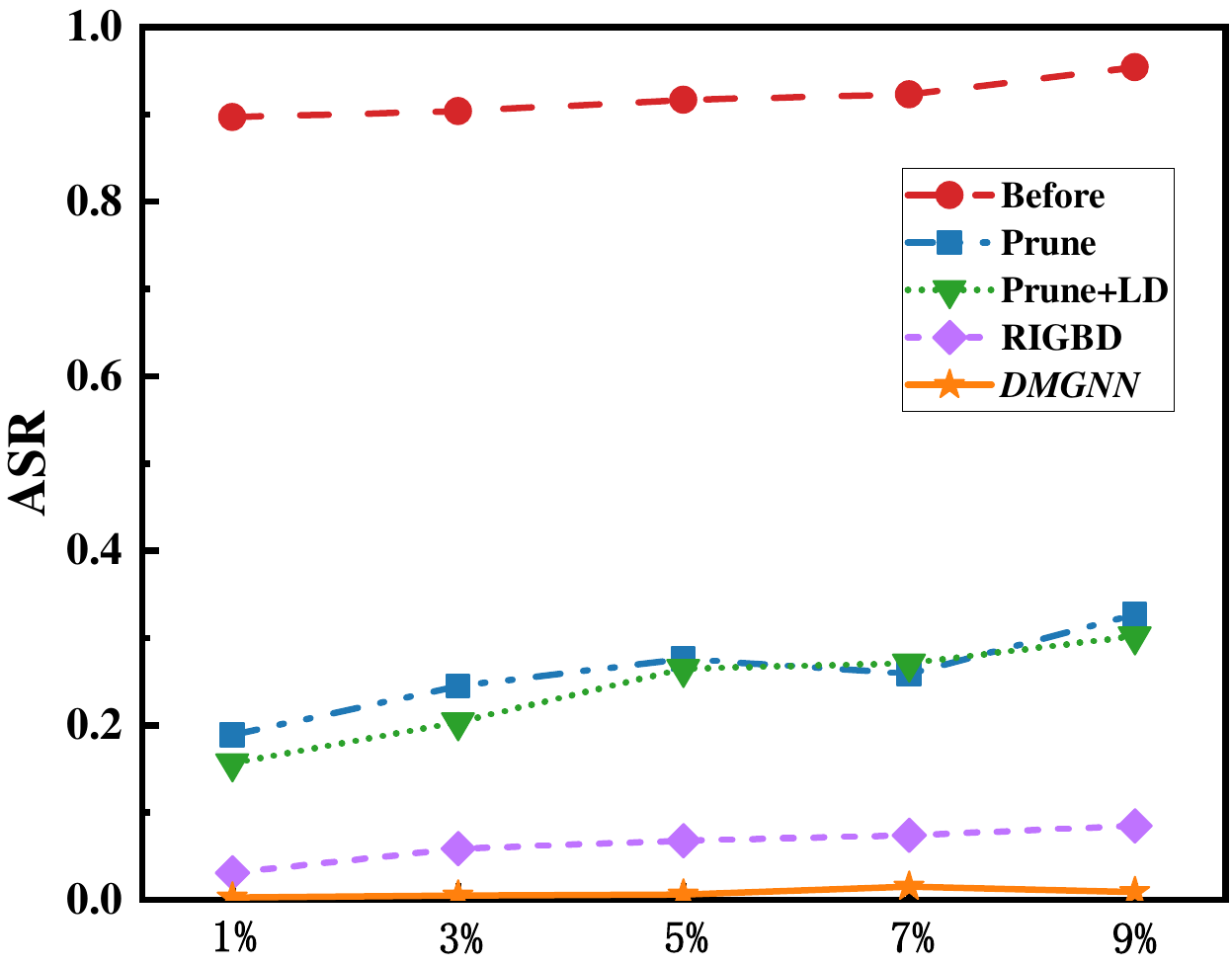}
			\label{F3-1}
	\end{minipage}}
	\subfigure[UGBA]{
		\begin{minipage}[t]{0.325\linewidth}
			\centering			\includegraphics[width=1\linewidth]{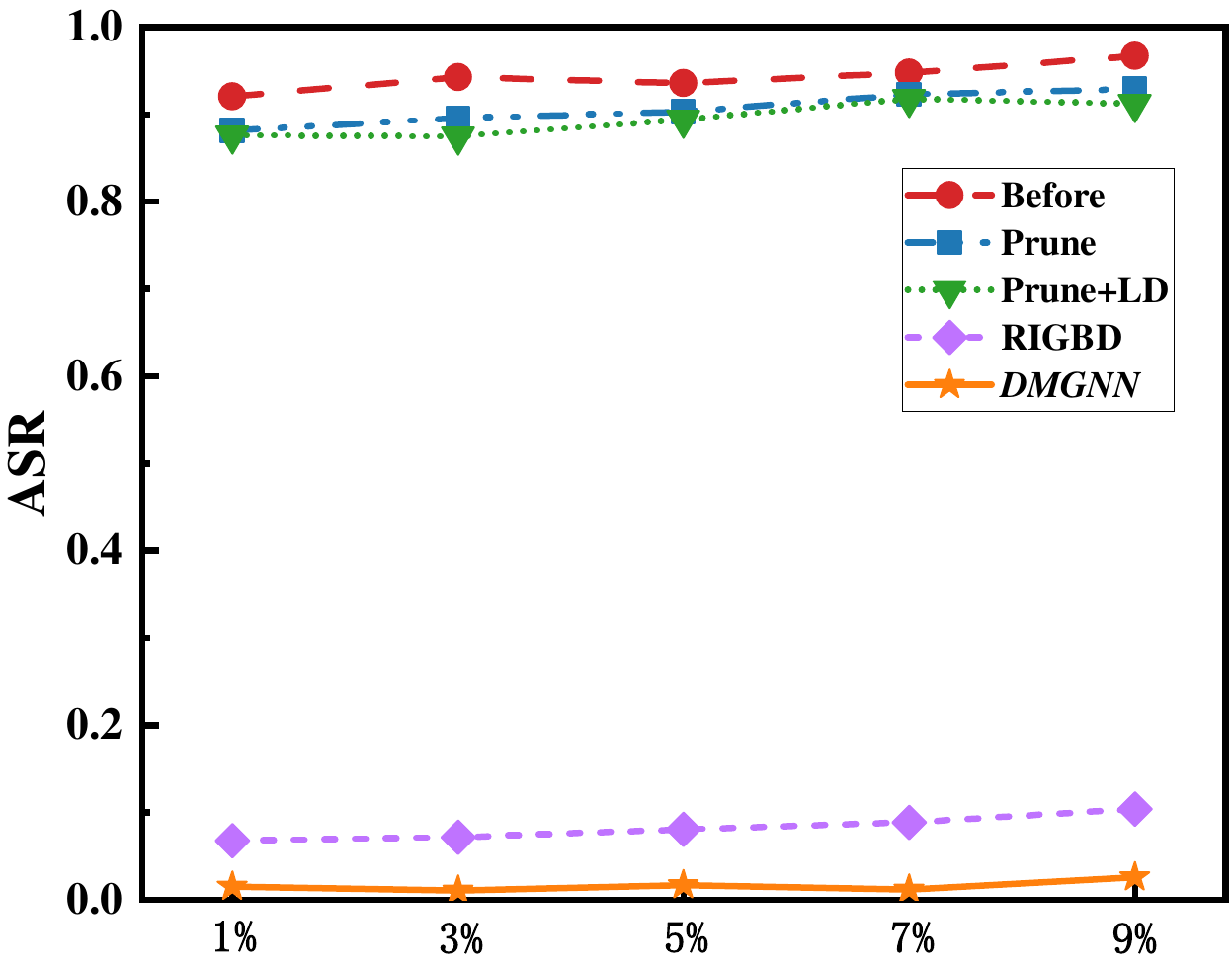}
			\label{F3-2}
	\end{minipage}} 
 \subfigure[DPGBA]{
		\begin{minipage}[t]{0.325\linewidth}
			\centering			\includegraphics[width=1\linewidth]{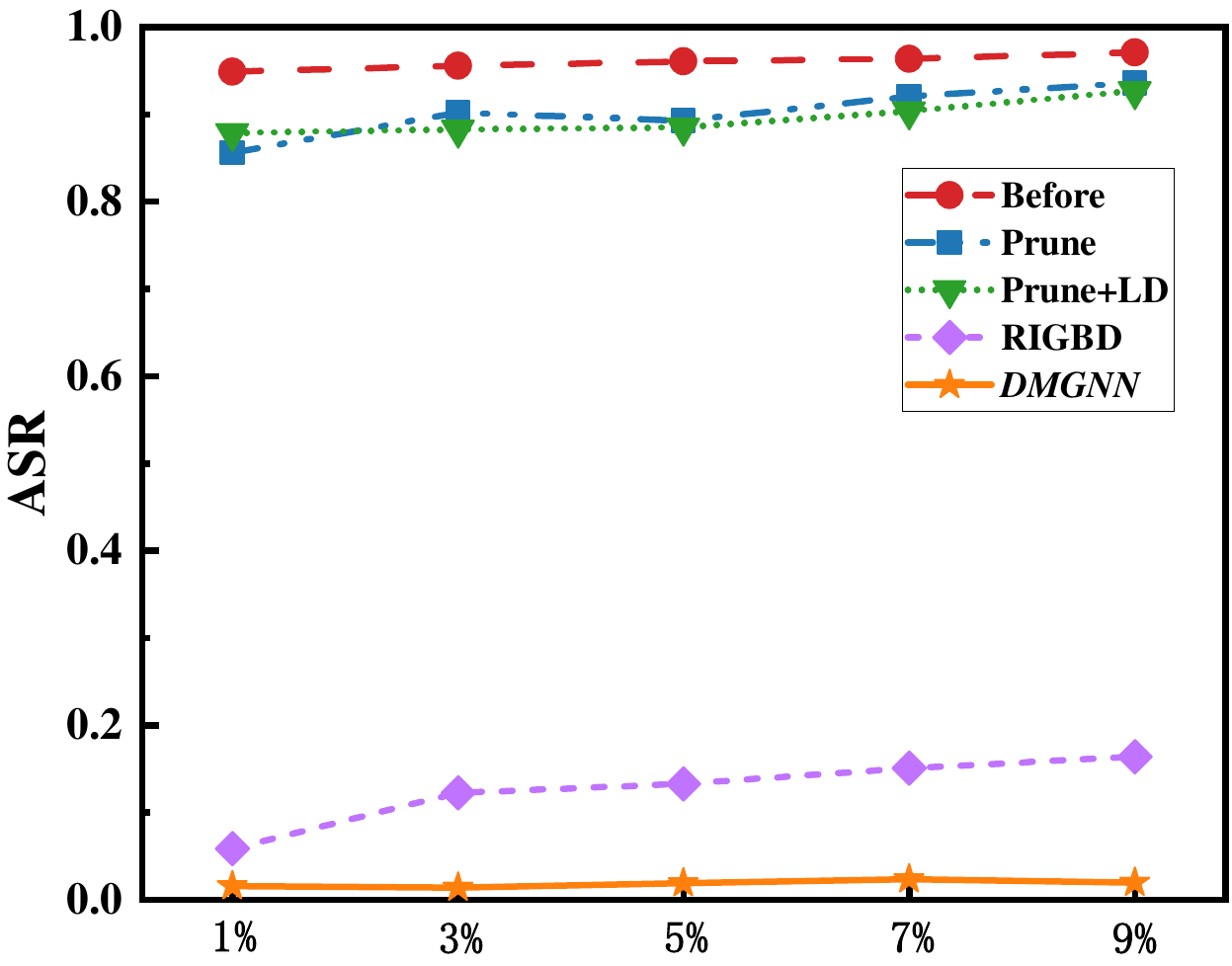}
			\label{F3-3}
	\end{minipage}}
\caption{Impacts of the different proportions of backdoored samples against three attacks on the Cora dataset.}
	\label{F3}
\end{figure*}

\begin{figure}
	\centering
	\subfigure[ASR]{
		\begin{minipage}[t]{0.485\linewidth}
			\centering			\includegraphics[width=1\linewidth]{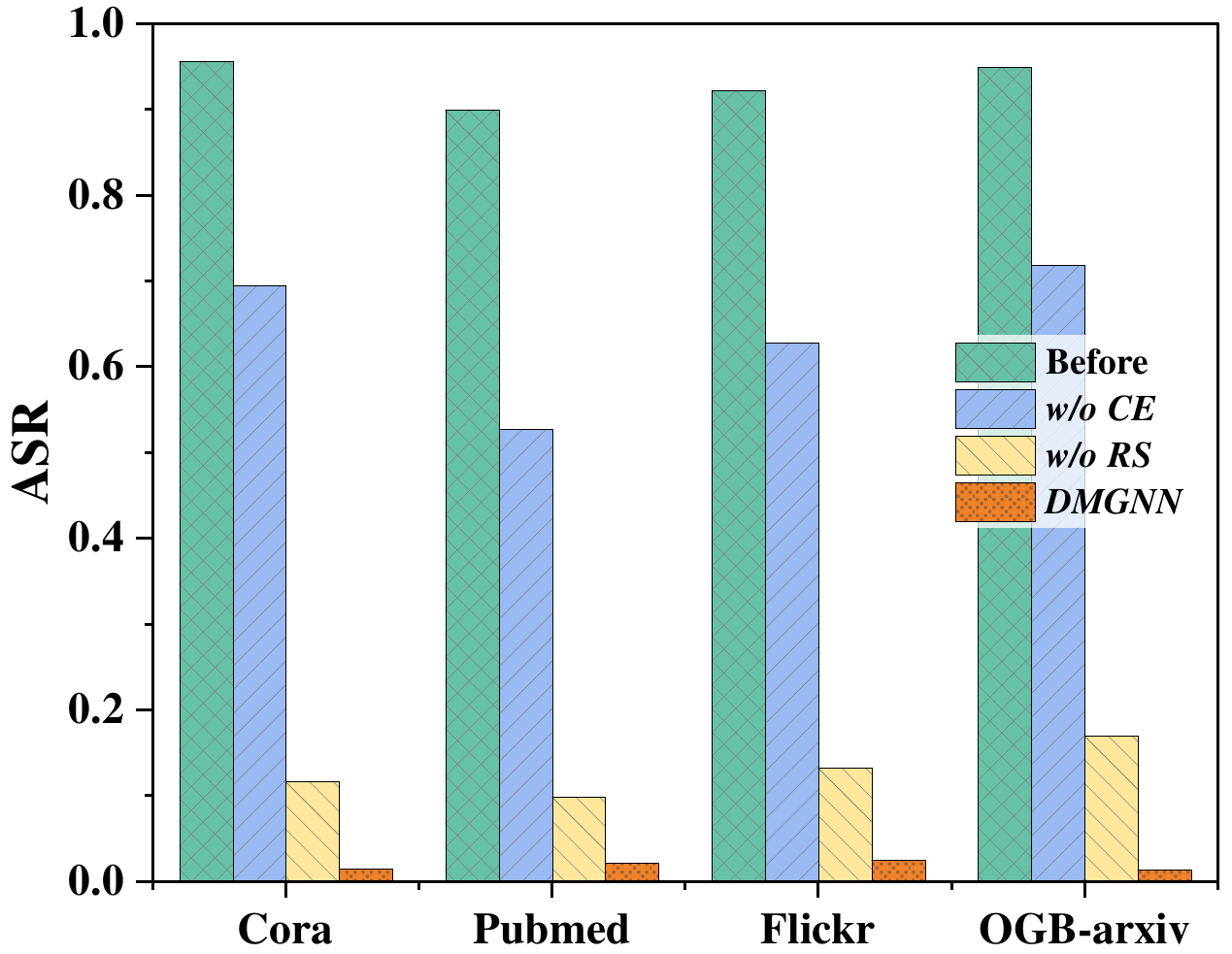}
			\label{F4-1}
	\end{minipage}}
	\subfigure[ACC]{
		\begin{minipage}[t]{0.485\linewidth}
			\centering			\includegraphics[width=1\linewidth]{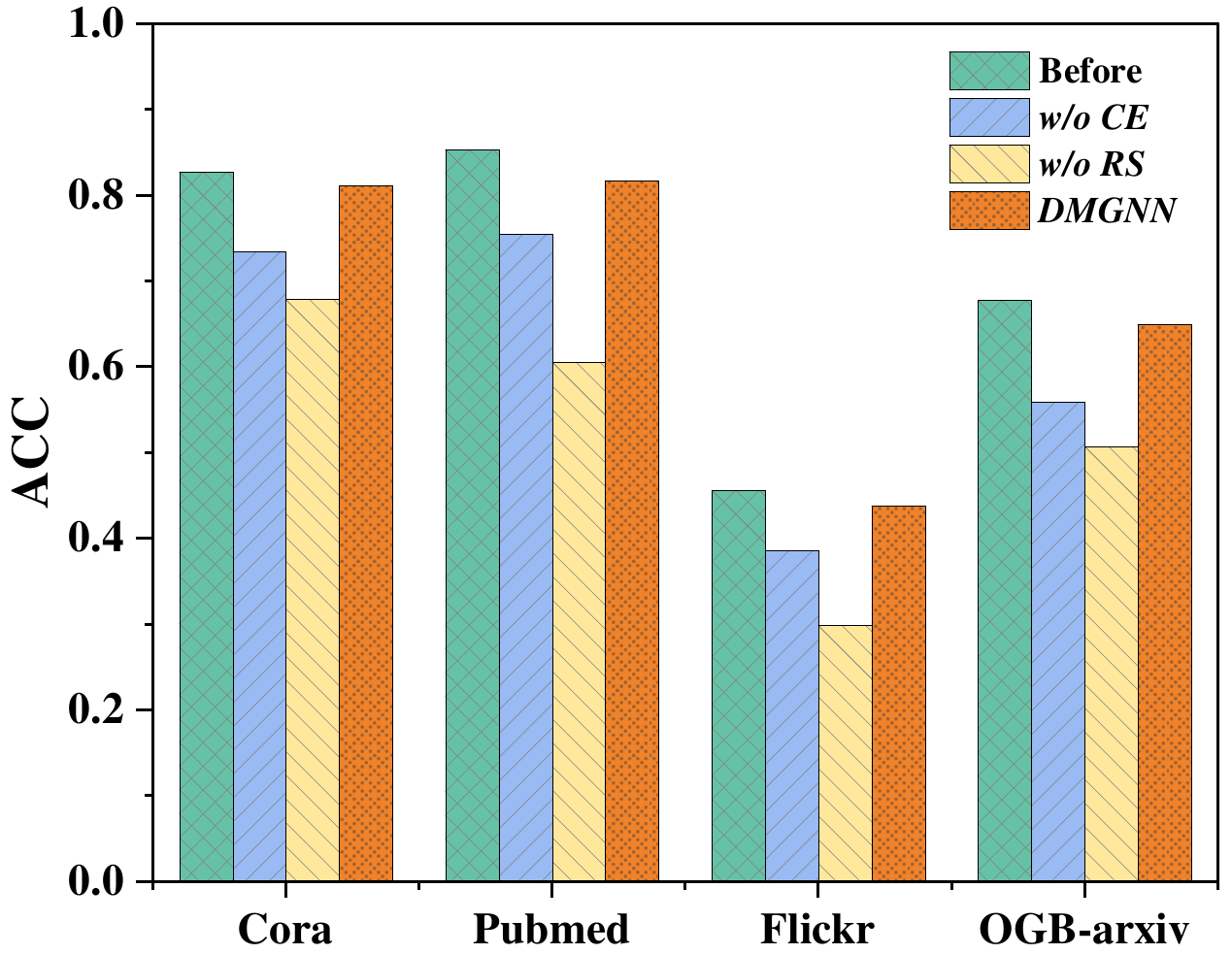}
			\label{F4-2}
	\end{minipage}} 
\caption{Ablation analysis against DPGBA attack on four real-world datasets.}
	\label{F4}
\end{figure}

\subsection{Further Understanding of \textit{DMGNN}  \Circled{\textbf{RQ2}}}
To further validate the performance of \textit{DMGNN}, we analyze the robustness of the defense of our method in the case of injecting different proportions of backdoor samples. Meanwhile, we also analyze the two hyper-parameters $\lambda$ and $p$ that affect the performance of the model, which are described in Section \ref{sec: appendix c} and Appendix \ref{sec: appendix d}.

\subsubsection{Impacts of the different proportions of backdoored samples.}
In our experiments, we set the number of injected backdoored features to 3\% of training data. We select the sizes of poisoned samples as \{1\%,3\%,5\%,7\%,9\%\} to further explore the effect of different numbers of injected backdoored features on \textit{DMGNN}. Our experiment selects the Cora dataset for validation on three SOTA attack methods and presents the experimental results in Fig. \ref{F3} and \ref{F6}. It is found from the experimental results that as the number of injected backdoored nodes increases, the ASR of \textit{DMGNN} keeps increasing, while the ACC of \textit{DMGNN} keeps decreasing. It is consistent with our expectation that the increase of backdoor nodes causes the model to learn more backdoor features, which enhances the effectiveness of the graph backdoor attack.

Despite the attack enhancement, \textit{DMGNN} can reduce the ASR to below 0.03. In Fig. \ref{F3-1}, the prune and prune+LD methods achieve good defense, while for UGBA and DPGBA attacks these two defense methods are essentially useless in Figure \ref{F3-2} and \ref{F3-3}. This is due to the fact that the GTA attack is an OOD backdoor attack, and it is able to distinguish the backdoor through the difference between trigger features and normal sample features by pruning. Thus, pruning can only defend against triggers with feature distributions that differ significantly from normal samples. However, for triggers with similar feature distributions with normal samples, the pruning method cannot accurately remove malicious samples, resulting in the model trained on randomly pruned data being ineffective in defending against backdoor attacks in GNNs.
In addition, the difference in the accuracy of \textit{DMGNN} compared to the accuracy before the defense can be controlled within 0.02.
Furthermore, the accuracy of \textit{DMGNN} can be maintained within a difference of 0.02 compared to the accuracy without defending. In Figure \ref{F5-1} and \ref{F5-3}, the accuracy of \textit{DMGNN} exceeds that of the three baseline defense methods. Among baseline methods, the RIGBD method shows the best performance, maintaining high accuracy under both UGBA and DPGBA attacks, as it is capable of defending against backdoor attacks with ID triggers.

\subsubsection{Impacts of the hyper-parameter $\lambda$}
\label{sec: appendix c}
As depicted in Figure \ref{F7}, the experimental outcomes reveal that the model performance variations when the hyper-parameter $\lambda$ is altered against the DPGBA attack. The hyper-parameter $\lambda$ controls the balance between distributional and counterfactual losses, which seriously impacts the defense efficacy of \textit{DMGNN}. In this paper, we set the hyper-parameter $\lambda$ to 0.1 and varied its values across \{0.005, 0.01, 0.05, 0.1, 0.5\} to
explore the impact of this hyper-parameter in our experiment.

When the hyper-parameter $\lambda$ is lower than 0.1, the ASR of \textit{DMGNN} gets increasingly higher as the value of $\lambda$ gets smaller. It is equivalent to the weakening of the contribution of counterfactual explanation, making it difficult for the model to defend against backdoor attacks with ID triggers. The reason is that it is not possible to accurately identify hidden triggers that are consistent with the normal sample feature distribution via distribution loss alone. When the hyper-parameter $\lambda$ is 0.5, the ACC of \textit{DMGNN} is lower than the value with $\lambda$ set to 0.1. On the Cora dataset, $\lambda$ of 0.5 is 2\% lower than $\lambda$ of 0.1. This is due to the increased contribution of the counterfactual explanation, which leads to weakening the properties of the same feature distribution of generated triggers. Meanwhile, the computational cost is excessively high for $\lambda$ of 0.5 despite the fact that the result is acceptable. Therefore, we adapt the value of $\lambda$ as 0.1 for our experiments.

\subsection{Ablation Analysis of \textit{DMGNN}  \Circled{\textbf{RQ3}}}
To clearly understand the role of each component of \textit{DMGNN}, we perform the ablation analysis. Our method employs the counterfactual explainable approach to identify potential ID or OOD triggers, so we eliminate counterfactual losses during training and name as \textit{w/o CE}. To demonstrate the validity of reverse sampling to further identify triggers, we eliminate the reverse sampling module and name as \textit{w/o RS}. The experimental results are shown in Figure \ref{F4}.

Under the DPGBA attack method, the attack success rate after \textit{w/o CE} defense is still high. Among them, the ASR of \textit{w/o CE} in the OGB-arxiv dataset is able to reach 0.718, indicating poor defense effectiveness. This shows that the counterfactual explainable method plays a key role in reducing the effectiveness of the graph backdoor attack. The ASR of \textit{w/o RS} can be significantly reduced, however, the accuracy is still worse than \textit{DMGNN}. For the Pubmed dataset, the ACC of \textit{w/o RS} is reduced by more than 20\% compared to \textit{DMGNN}. Therefore the reverse sampling model is crucial for the model performance.

\section{SUMMARY AND FUTURE WORK}
\label{sec: SUMMARY AND FUTURE WORK}
In this paper, we present \textit{DMGNN} a novel detect and mitigate method against various types of graph backdoor attacks with ID and OOD triggers. By adding noise and denoising variations in the backdoored sample prediction results, \textit{DMGNN} can effectively identify the hidden ID and OOD triggers against graph backdoor attacks. Considering the diversity of the generated explainable graphs and the fact that it is still impossible to eliminate the influence of triggers, we propose the reverse sampling pruning method to filter and remove the triggers precisely. \textit{DMGNN} can successfully eliminate the stealthiness to improve the defense effectiveness and the model performance against backdoor attacks. In different real-world datasets, we validate extensive experimental demonstrations that \textit{DMGNN} outperforms every baseline method under the different SOTA backdoor attack methods in GNNs. In future work, we will further explore how to mitigate the latest backdoor attacks in graph classification tasks. Simultaneously, GNNs in federated learning scenarios have also suffered from backdoor attacks, we continue to investigate effective defense mechanisms.


\newpage
\bibliographystyle{ACM-Reference-Format}
\bibliography{dmgnn}

\appendix

\begin{table}[ht]\small
	\centering
 \caption{\textbf{Analysis of the datasets}
 \label{table: Analysis of the datasets}}
\renewcommand{\arraystretch}{1.5} 
 \setlength{\tabcolsep}{2mm}{
\begin{tabular}{ccccccc}
\toprule[1pt]
\toprule[0.5pt]
\textbf{Datasets} & \textbf{\#Nodes} & \textbf{\#Edges} & 
\textbf{\#Feature} & \textbf{\#Classes} &  \\ \hline
Cora & 2,708 & 5,429 
& 1,433 & 7\\
Pubmed & 19,717 & 44,338 
& 500 & 3 \\
Flickr & 89,250 & 899,756 
& 500 & 7 \\
OGB-arxiv & 169,343 & 1,166,243 
& 128 & 40 \\   
\toprule[0.5pt]
\toprule[1pt]
\end{tabular}
}
\end{table}

\begin{algorithm}[t]
\caption{Counterfactual
Explanation} 
	\label{A1}
    \KwIn{Original graph set $G_0$, Graph dataset set $Dataset=\{(v_1,y_1),(v_2,y_2),...,(v_\mathcal{N},y_\mathcal{N})\}$, Diffusion process $q(\cdot)$, Denoising model $p(\cdot)$, Noise level $n$, Hyper-parameter $\lambda$, Backdoored model $\mathcal{G}_d$, Label space $\mathbb{L}$, Remove sample process $r(\cdot)$.} 
    Initialize model parameters for $p(\cdot)$\;    
    \For{$y_p\in\mathbb{L}$}{
    \While{not converged yet}
    {
    \texttt{//Adding
noise}\;
	\For{$n = 1$ to $N$}
	{ 
        Generate noise graph $G_n^{y_p}$ via Eq(\ref{Eq3});
	}
 \texttt{//Removing noise}\;
    \For{$n = N$ to $0$}
	{  
        Generate counterfactual explanation graph $\overline{G}_n^{y_p} \leftarrow p(G_0^{y_p}|G_N^{y_p})$;
	}
    Computer the counterfactual loss $\mathcal{L}_{c}$ via Eq(\ref{Eq4})\;       
    Computer the distribution loss $\mathcal{L}_{d}$ via Eq(\ref{Eq5})\;
    $\mathcal{L} \leftarrow \lambda\mathcal{L}_{c} + \mathcal{L}_{d}$\;
    Update model parameters for $p(\cdot)$ to min the loss $\mathcal{L}$\;
    }    \If{$\mathcal{G}_d(r(v^{y_p},\overline{G}_0^{y_p}))$ != $y_p$}{
    \KwOut{Counterfactual
explanation graph $\overline{G}_0^{y_p}$.}}
    }
\end{algorithm}

\begin{figure*}
	\centering
	\subfigure[GTA]{
		\begin{minipage}[t]{0.325\linewidth}
			\centering	\includegraphics[width=1\linewidth]{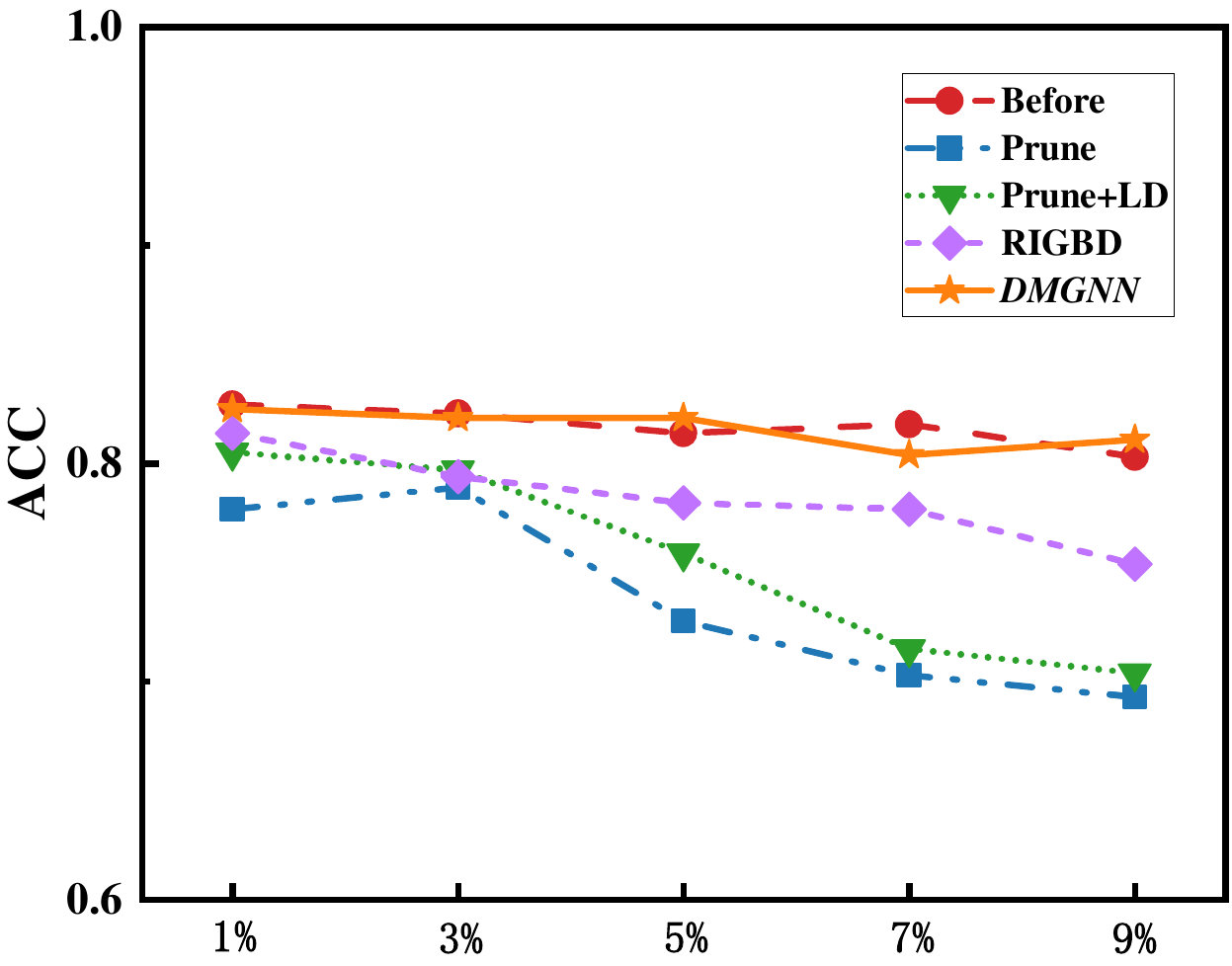}
			\label{F5-1}
	\end{minipage}}
	\subfigure[UGBA]{
		\begin{minipage}[t]{0.325\linewidth}
			\centering			\includegraphics[width=1\linewidth]{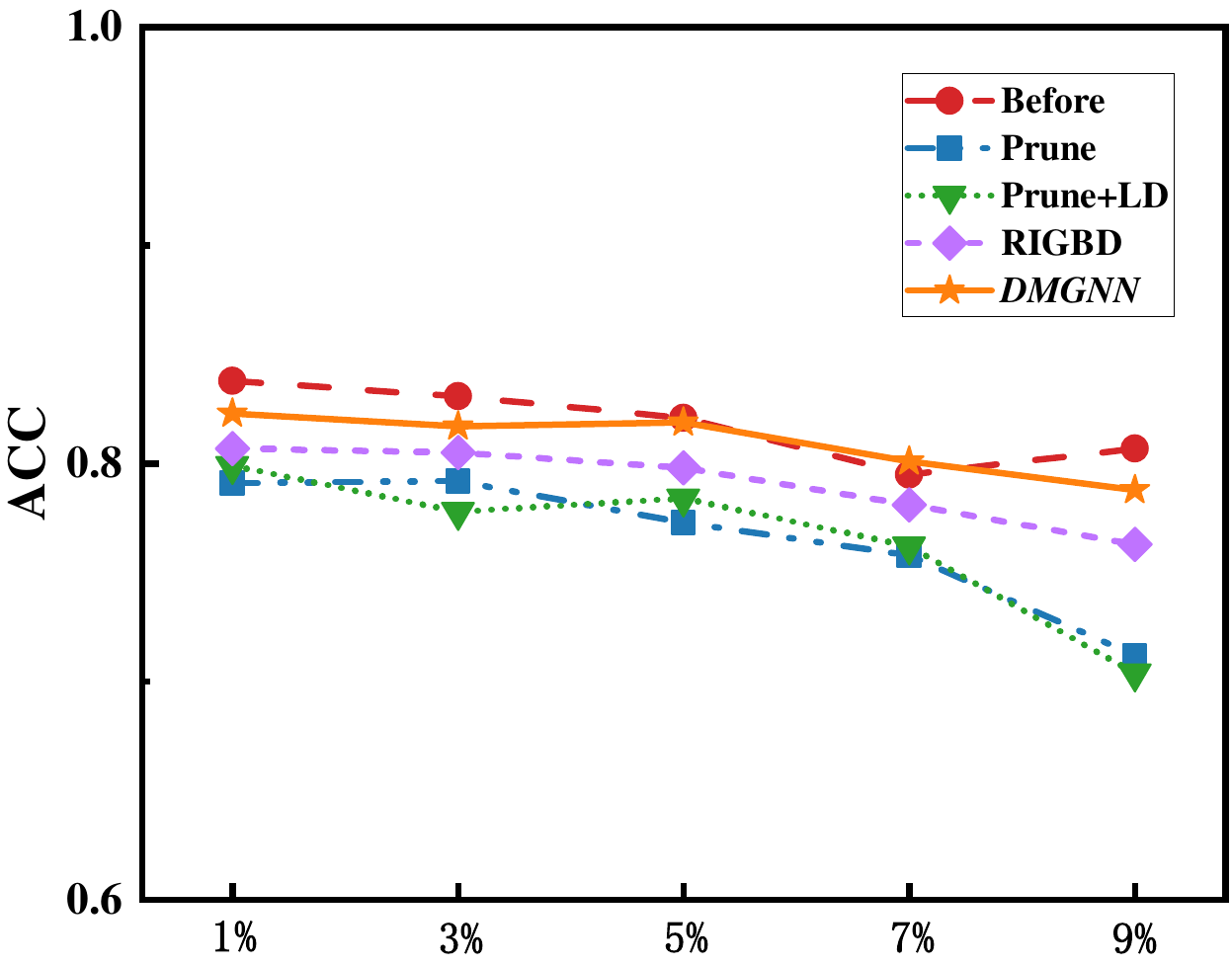}
			\label{F5-2}
	\end{minipage}} 
 \subfigure[DPGBA]{
		\begin{minipage}[t]{0.325\linewidth}
			\centering			\includegraphics[width=1\linewidth]{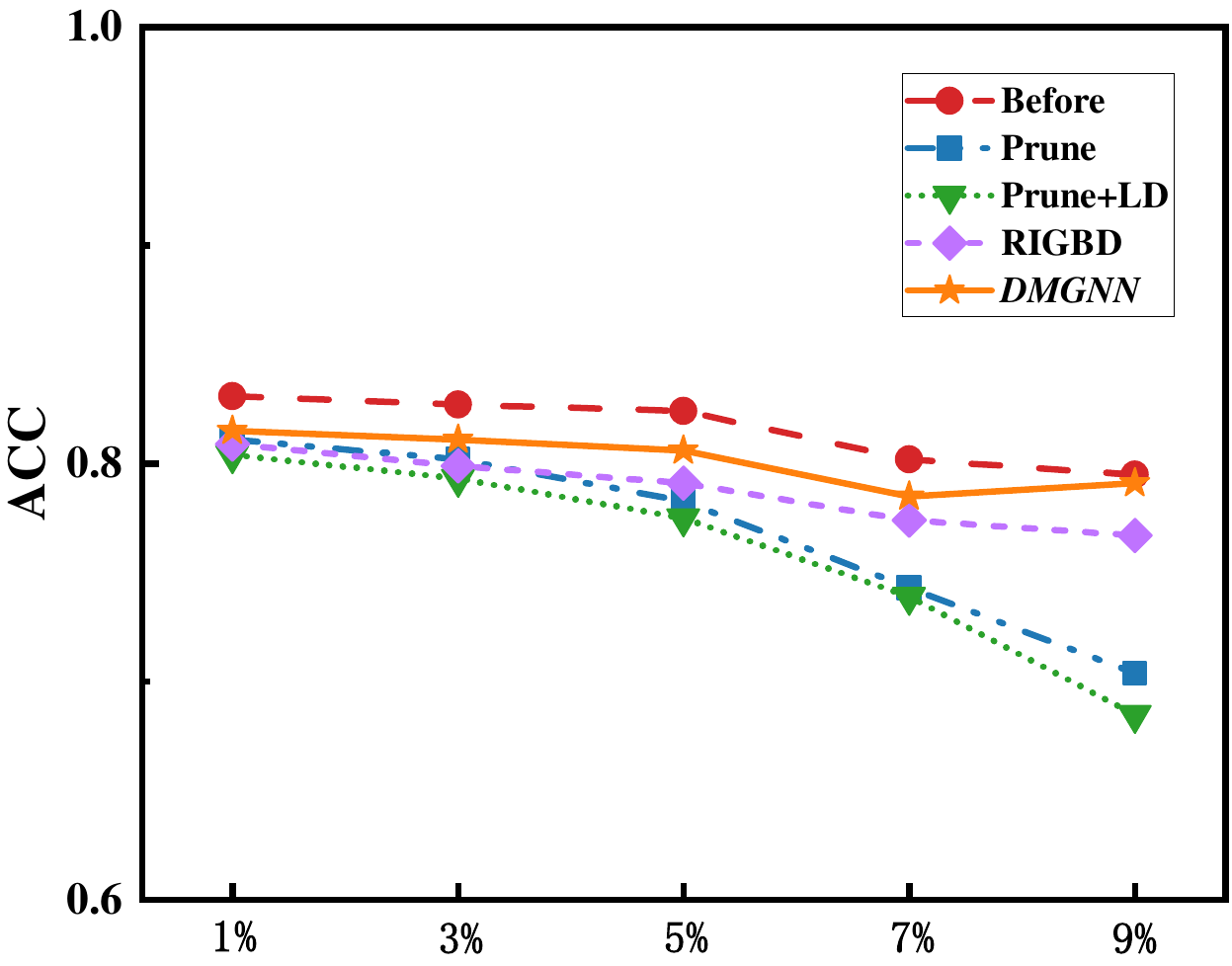}
			\label{F5-3}
	\end{minipage}}
\caption{Impacts of the different proportions of backdoored samples against three attacks on the Cora dataset.}
	\label{F5}
\end{figure*}

\begin{figure*}
	\centering
	\subfigure[ASR with GTA method]{
		\begin{minipage}[t]{0.32\linewidth}	\centering	\includegraphics[width=1\linewidth]{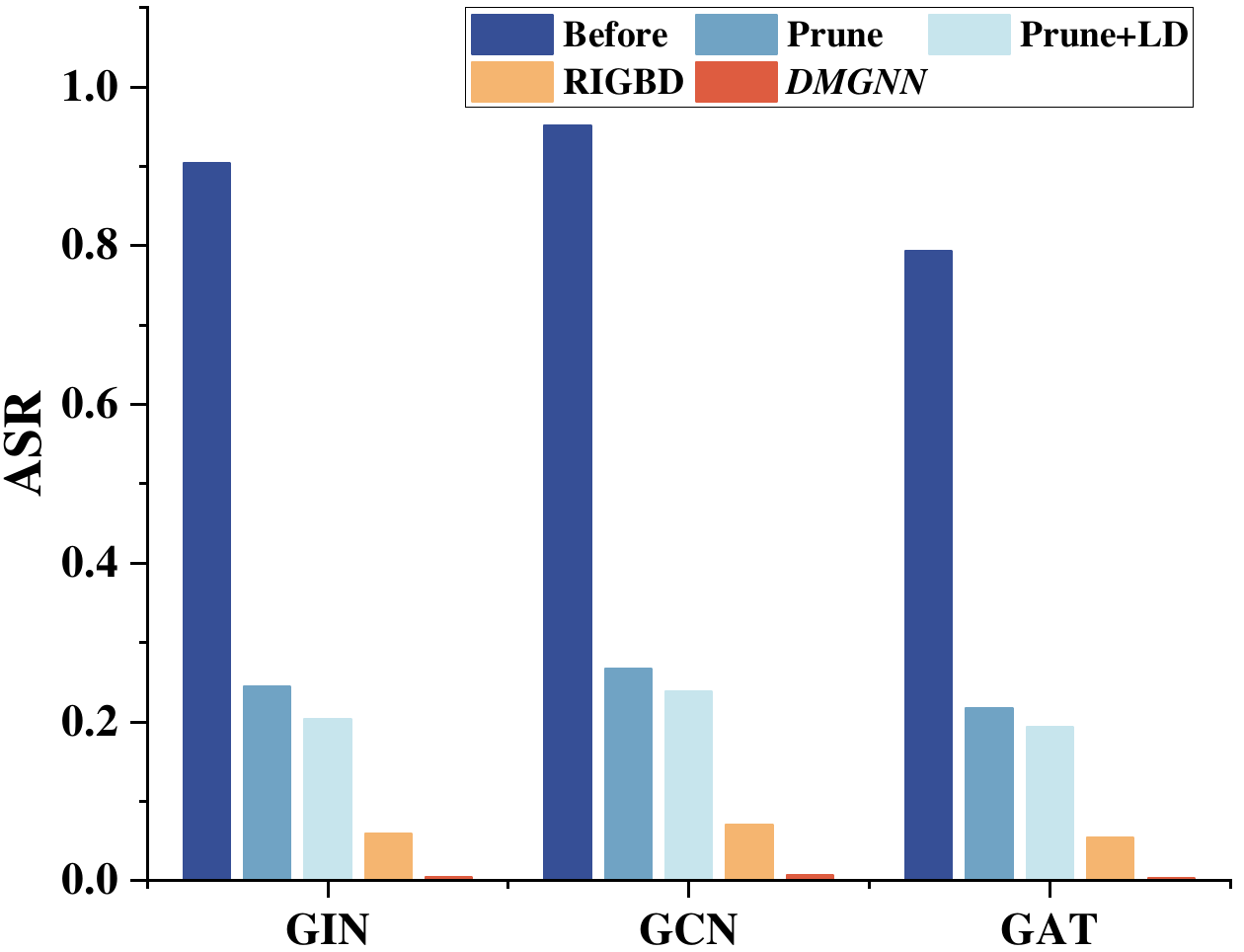}
			\label{F6-1}
	\end{minipage}}
	\subfigure[ASR with UGBA method]{
		\begin{minipage}[t]{0.32\linewidth}	\centering	\includegraphics[width=1\linewidth]{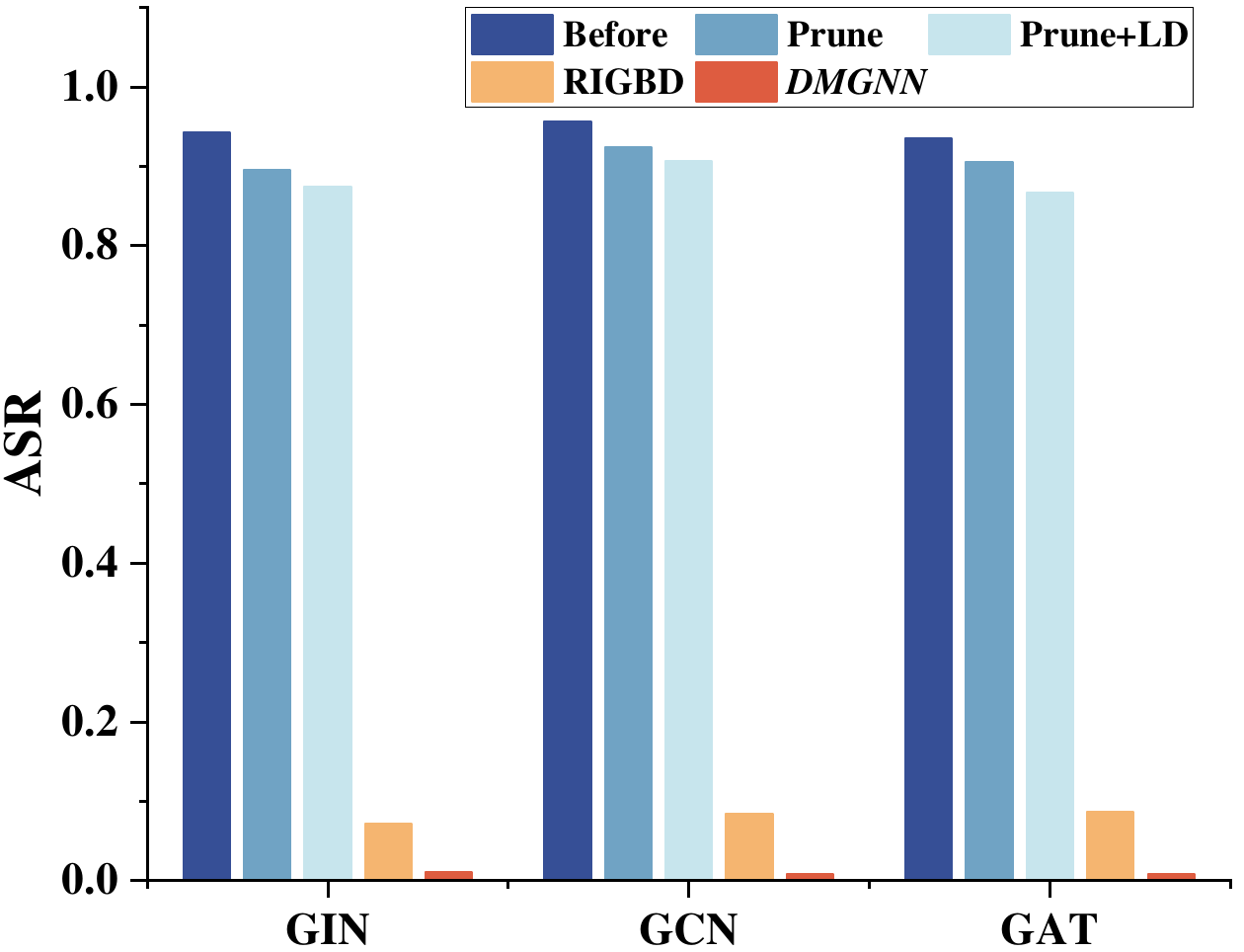}
			\label{F6-2}
	\end{minipage}} 
 \subfigure[ASR with DPGBA  method]{
		\begin{minipage}[t]{0.32\linewidth}
			\centering	\includegraphics[width=1\linewidth]{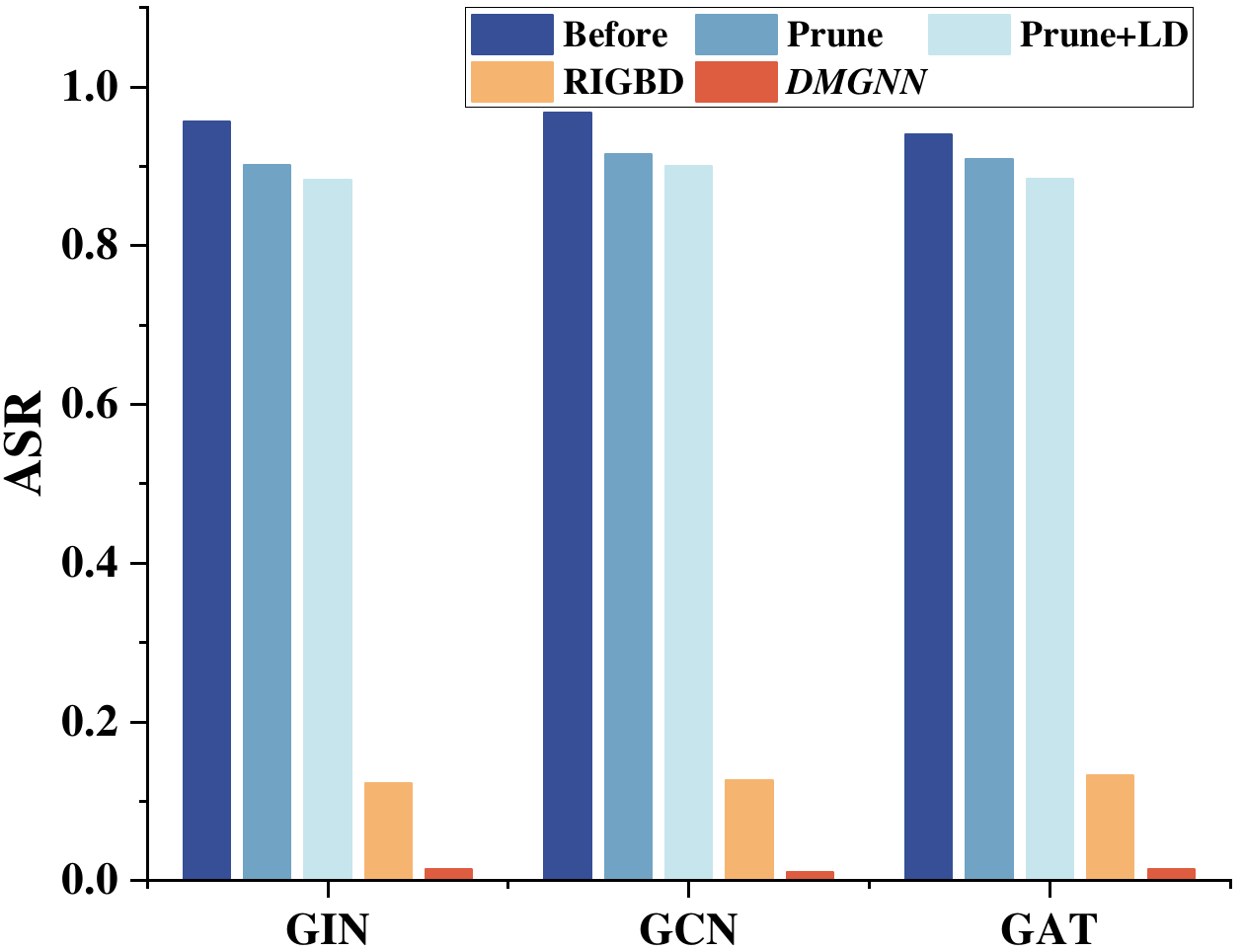}
			\label{F6-3}
	\end{minipage}}
 
 	\subfigure[ACC with GTA method]{
		\begin{minipage}[t]{0.32\linewidth}
			\centering	\includegraphics[width=1\linewidth]{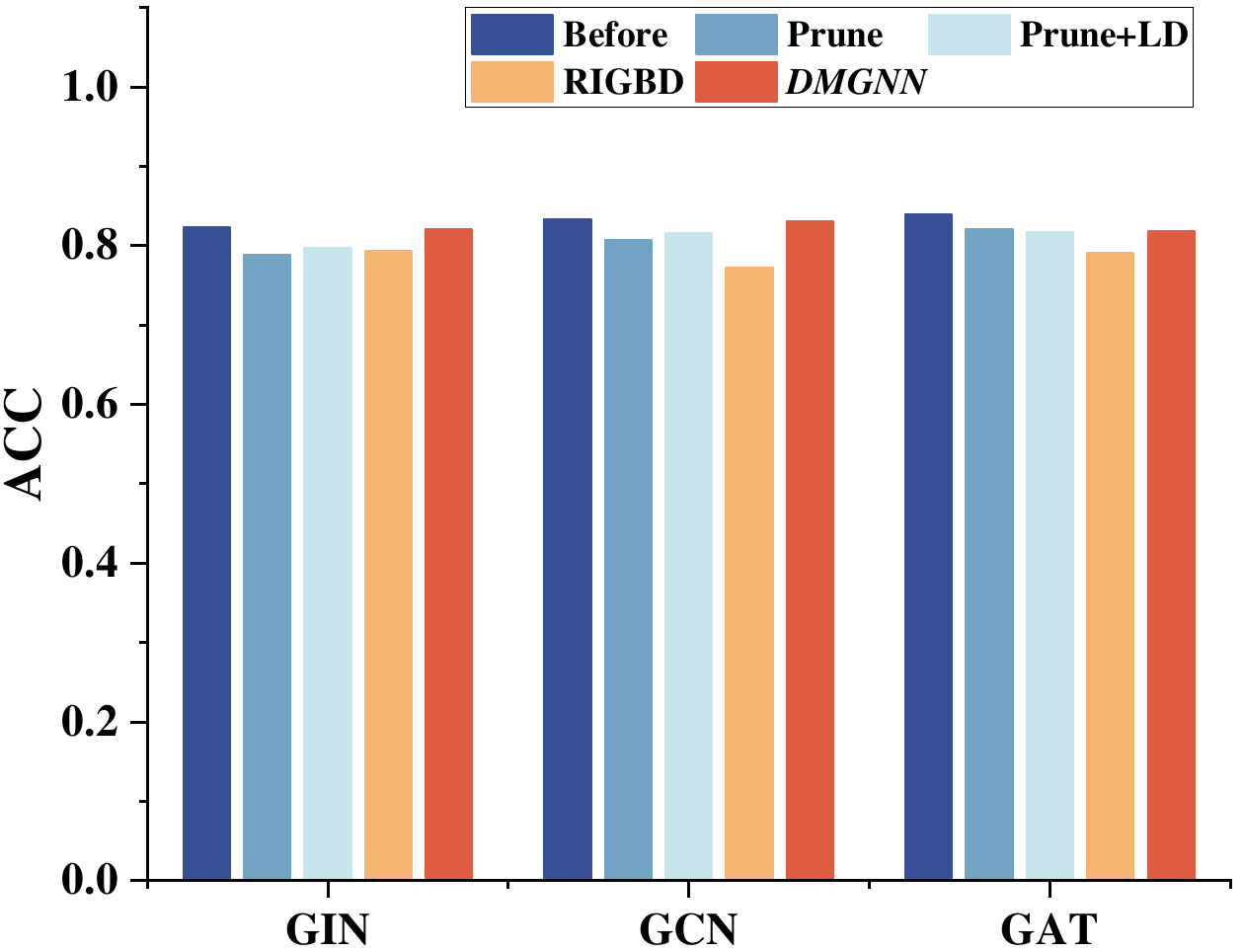}
			\label{F6-4}
	\end{minipage}}
 \subfigure[ACC with UGBA method]{
		\begin{minipage}[t]{0.32\linewidth}
			\centering	\includegraphics[width=1\linewidth]{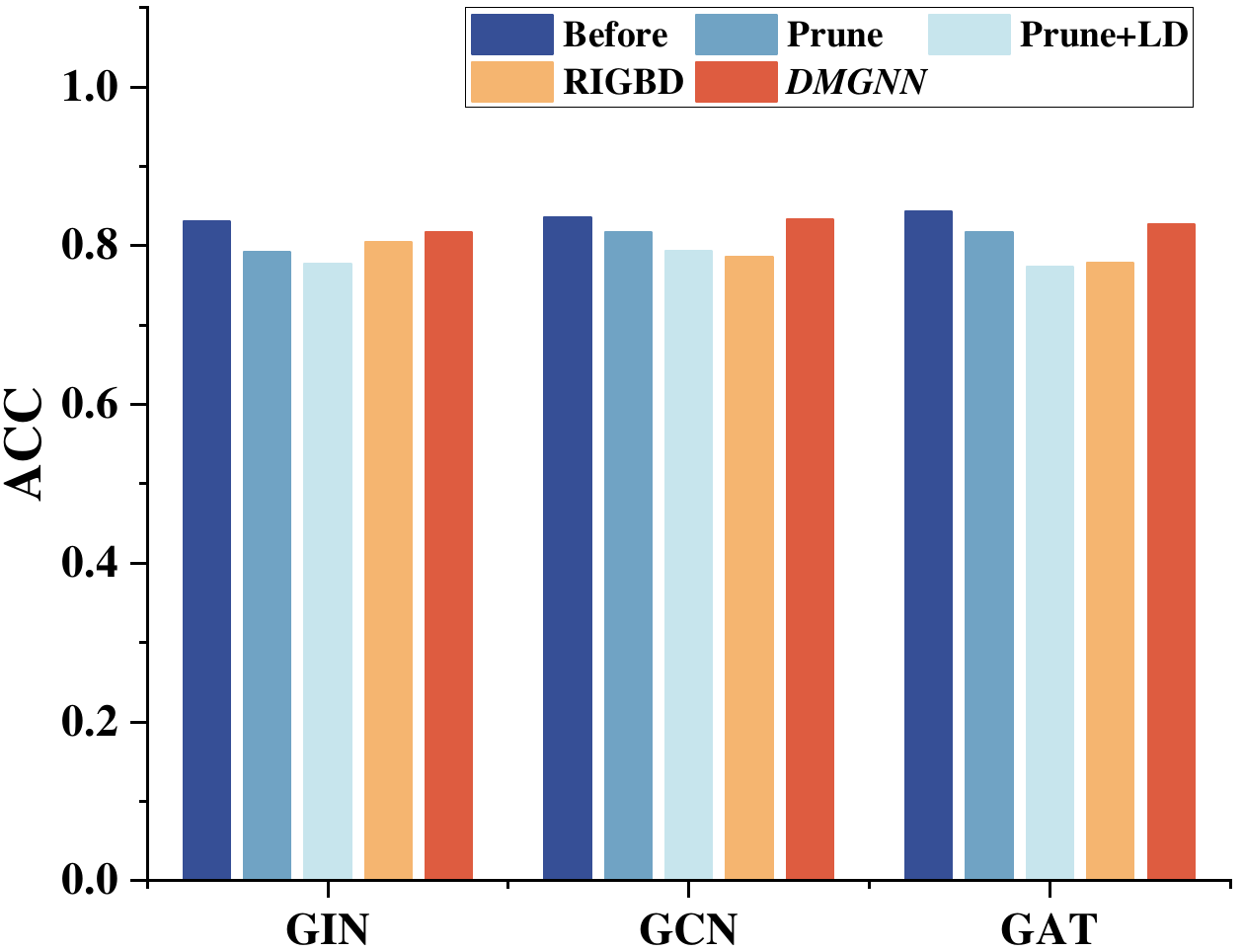}
			\label{F6-5}
	\end{minipage}}
 \subfigure[ACC with DPGBA method]{
		\begin{minipage}[t]{0.32\linewidth}
			\centering	\includegraphics[width=1\linewidth]{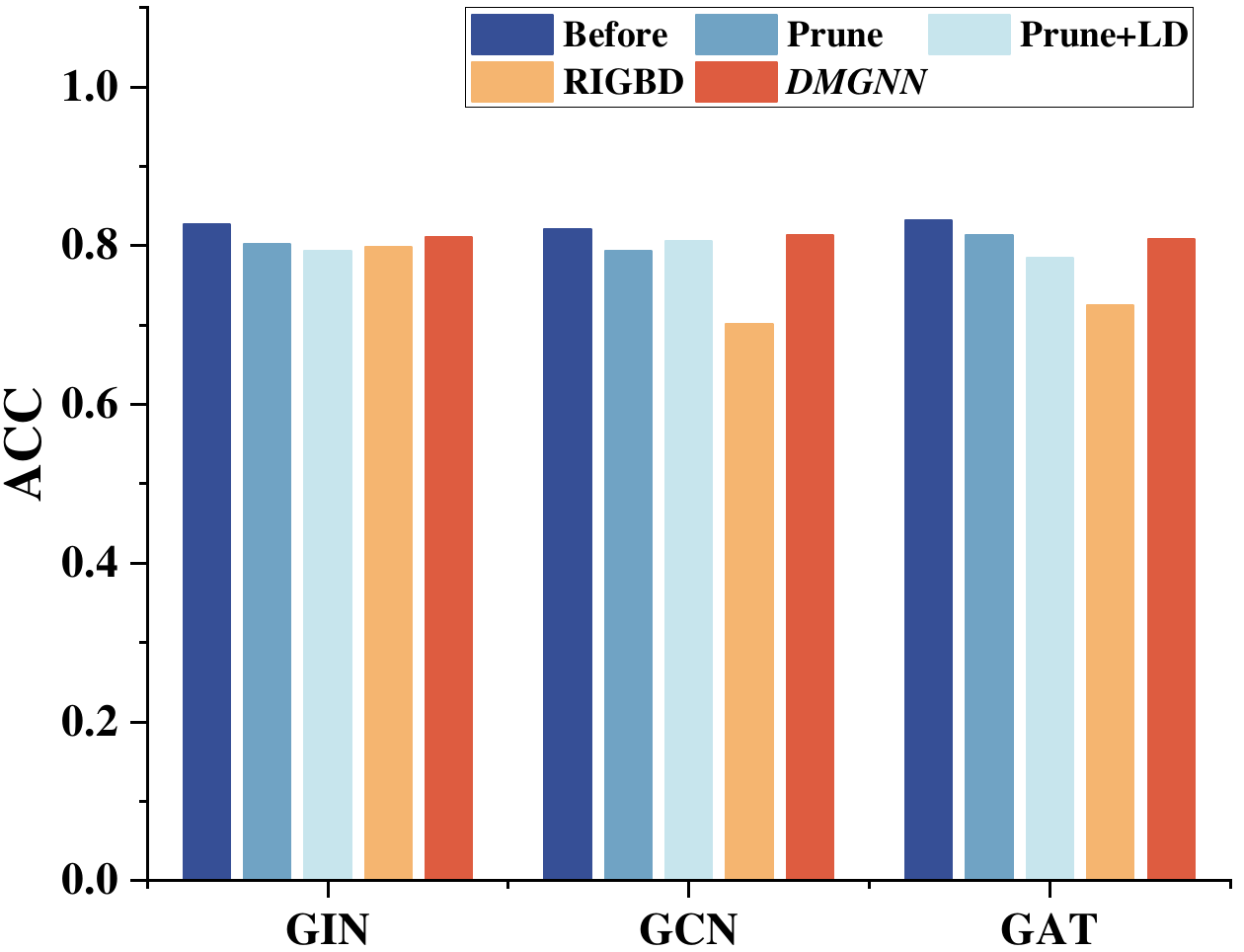}
			\label{F6-6}
	\end{minipage}}
 \caption{Impacts of model types against three attacks on the Cora dataset.}
	\label{F6}
\end{figure*}

\begin{figure}
	\centering
	\subfigure[ASR]{
		\begin{minipage}[t]{0.485\linewidth}
			\centering	\includegraphics[width=1\linewidth]{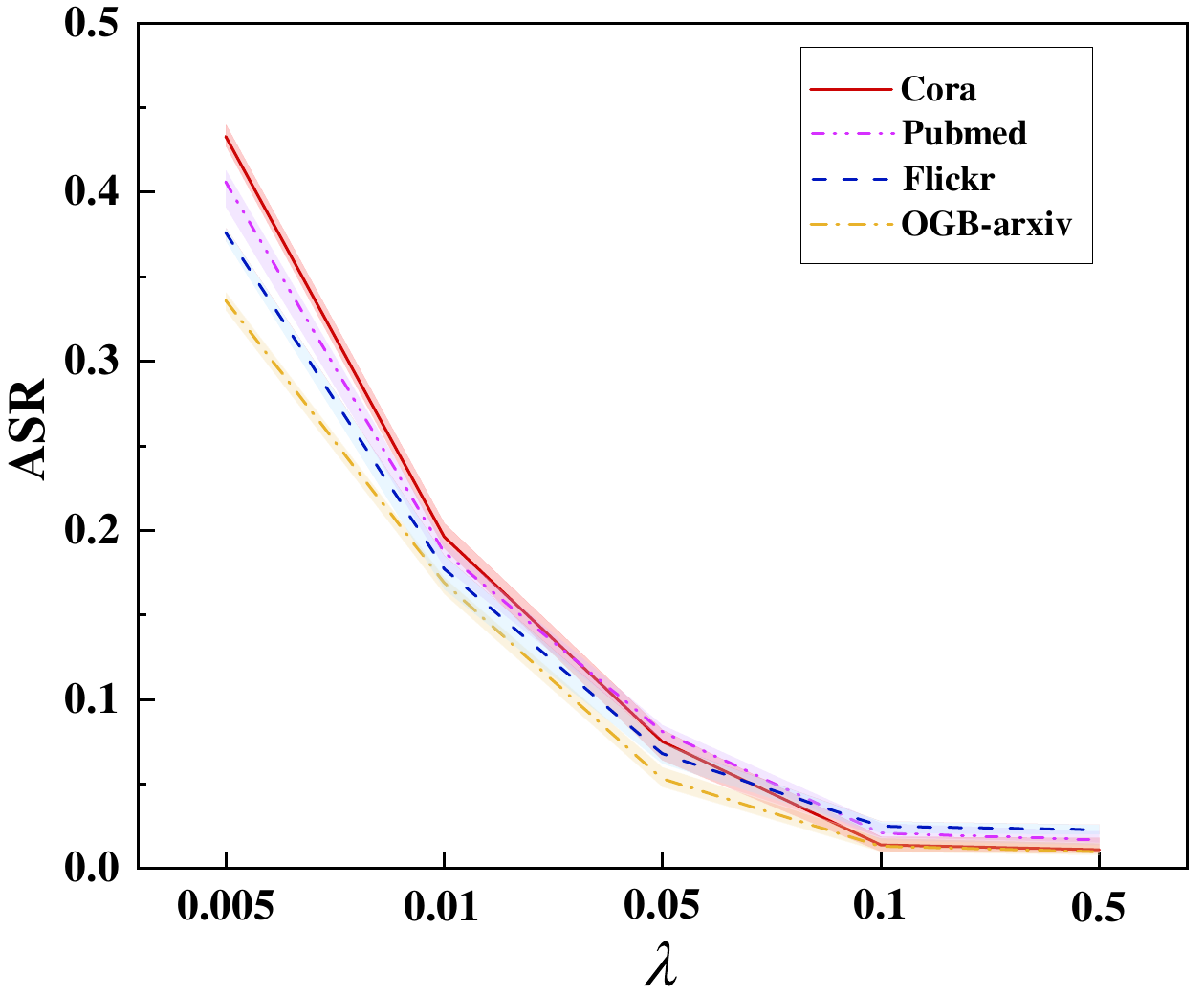}
			\label{F7-1}
	\end{minipage}}
	\subfigure[ACC]{
		\begin{minipage}[t]{0.485\linewidth}
			\centering	\includegraphics[width=1\linewidth]{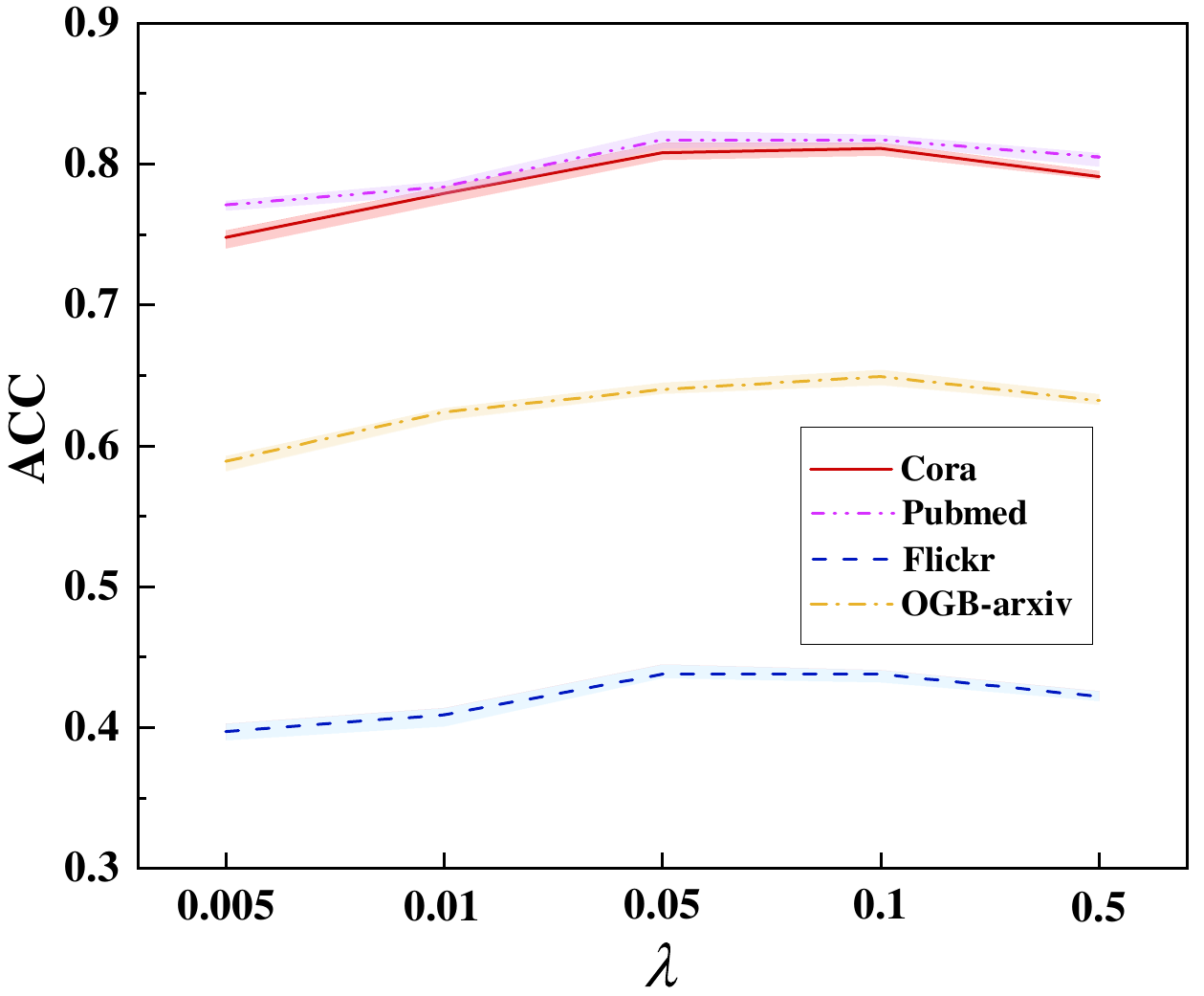}
			\label{F7-2}
	\end{minipage}} 
\caption{$\lambda$ analysis against DPGBA attack on four datasets.}
	\label{F7}
\end{figure}

\begin{figure}
	\centering
	\subfigure[ASR]{
		\begin{minipage}[t]{0.485\linewidth}
			\centering			\includegraphics[width=1\linewidth]{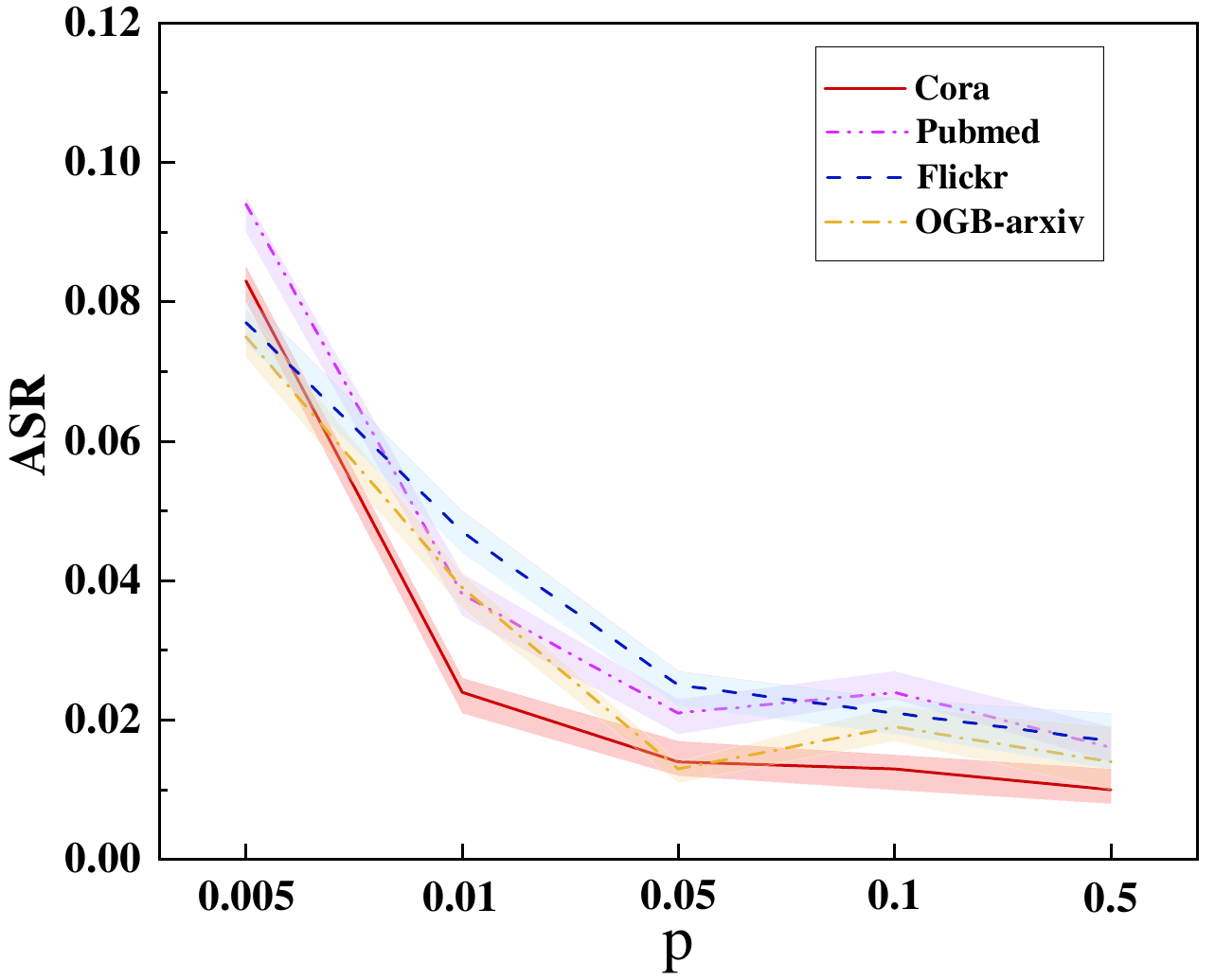}
			\label{F8-1}
	\end{minipage}}
	\subfigure[ACC]{
		\begin{minipage}[t]{0.485\linewidth}
			\centering			\includegraphics[width=1\linewidth]{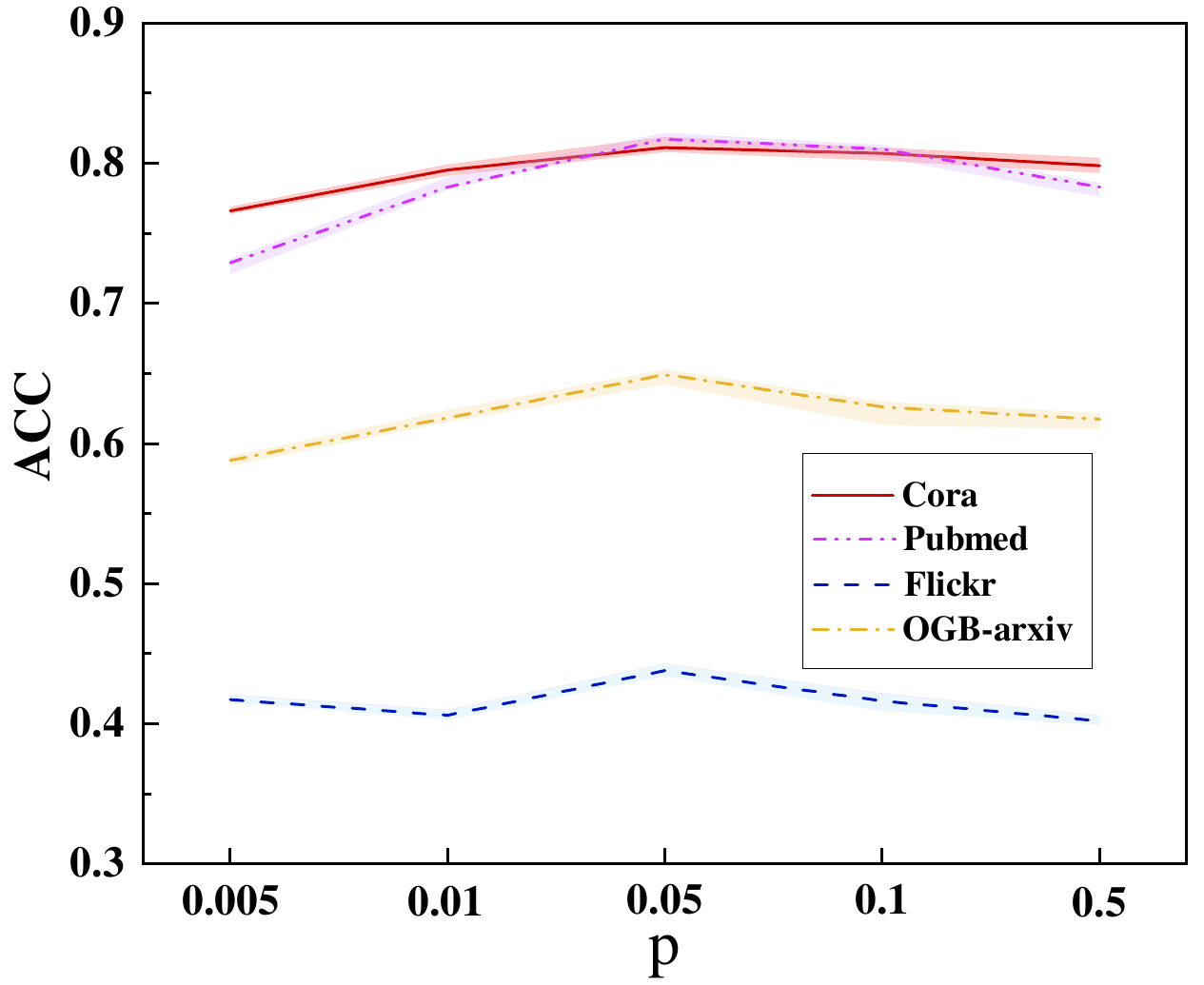}
			\label{F8-2}
	\end{minipage}} 
\caption{$p$ analysis against DPGBA attack on four datasets.}
	\label{F8}
\end{figure}

\section{Details on Datasets}
\label{sec: Details on Datasets}
\begin{itemize}
    \item \textbf{Cora:} 
    It is a citation network dataset commonly used for evaluating GNN models, where nodes represent papers and edges represent citation relationships between them. The dataset consists of 2,708 nodes and 5,429 edges. Each node is associated with a feature vector of 1,433 dimensions.
	
    \item \textbf{Pubmed:} It is another widely used citation network dataset, derived from biomedical research papers. In this dataset, nodes correspond to documents, and edges represent citation links. The dataset consists of 19,717 nodes and 44338 edges, which is larger than the Cora dataset.
    
    \item \textbf{Flickr:} It is a social network dataset where nodes represent users of the Flickr photo-sharing platform, and edges denote interactions between them, such as comments or likes. The dataset includes 89,250 nodes and has seven classes. Thus, it is a 7-class dataset in the node classification task.

    \item \textbf{OGB-arxiv:} The dataset represents a large-scale citation network, where nodes are ArXiv papers and edges indicate citation relationships. The dataset includes 169,343 nodes and 1,166,243 edges. A 128-dimensional feature vector represents each paper. The dataset is categorized into 40 classes and the task is to predict the subject area of each paper.
\end{itemize}

\section{Impacts of model types}
\label{sec: Different types of GNNs}
Figure \ref{F6} illustrates the impact of model types on \textit{DMGNN} against different attack methods. We validate our method on three widely recognized GNN models, including GIN, GCN, and GAT. Despite encountering various forms of attacks and complex models, \textit{DMGNN} consistently proves effective within acceptable margins. It is noteworthy that the ASR for the three models consistently maintains a minimum value of less than 2\% when subjected to different attack scenarios. Moreover, the accuracy of \textit{DMGNN} exceeds 80\% under different models, which is superior to all baseline methods. This finding emphasizes that the defense effectiveness of our approach is independent of model type.

\section{Impacts of the hyper-parameter $p$}
\label{sec: appendix d}
In our experiment, $p$ serves as a tunable parameter in the confidence of explanations, controlling the trigger after reverse sampling. We set $p$ to 0.9 and varied its values across \{0.8, 0.85, 0.9, 0.95, 1\} to explore the impact of this hyper-parameter. Figure \ref{F8} illustrates the effect of the hyper-parameter $p$ on \textit{DMGNN} against the DPGBA attack under four different datasets.

It can be observed that, overall, as the hyper-parameter $p$ decreases, the ASR of our method consistently increases. This indicates that when the explainable confidence is low, the obtained ID explainable graph becomes less important and thus is not necessarily the hidden trigger set by the attacker. It will cause the \textit{DMGNN} defense performance to degrade.  Conversely, as $p$ increases, the ASR of our method gradually decreases. When $p$ reaches 1, the ASR of \textit{DMGNN} is at its lowest. However, the ASR of \textit{DMGNN} at $p$ = 1 is not as low as when $p$ = 0.9. This is because, when the explanation confidence is set to 1, only a single explanation graph is generated, preventing optimal selection of the trigger based on prediction results. Taking all factors into account, we select $p$ = 0.9 as the hyper-parameter for our experiments.

\end{document}